# Conditions for Altermagnetism in $RuO_2$

Himanshu Mavani, Zhonglin He, and Evgeny Y. Tsymbal[†]

*Department of Physics and Astronomy & Nebraska Center for Materials and Nanoscience, University of Nebraska, Lincoln, Nebraska 68588-0299, USA*

The magnetic ground state of rutile $RuO_2$ remains disputed, raising a central question: under what structural and chemical conditions can this material support altermagnetism? We address this question by combining a critical assessment of experimental and theoretical studies with first-principles calculations of epitaxial strain, point defects, doping, and interfaces. Bulk-sensitive measurements reported in the literature are consistent with a nonmagnetic ground state, while magnetic signatures in thin films depend strongly on their structural environment and do not necessarily establish homogeneous altermagnetic order. Our calculations identify substrate-imposed epitaxial strain as the leading factor favoring altermagnetism among the conditions examined and predict several substrate-orientation combinations that stabilize an altermagnetic ground state. Carrier doping and chemical substitution modify this strain-induced magnetism in distinct ways. Uniform hole doping strengthens the magnetic tendency, whereas Ru vacancies provide little benefit and can weaken the magnetic state through local structural distortions. Cr or Mn substitution, by contrast, enhances Ru magnetic moments and further stabilizes the strained altermagnetic phase. These results distinguish the effects of carrier doping from those of local structural and chemical changes and identify concrete routes to realizing altermagnetism in $RuO_2$ films.

[†] *E-mail*: tsymbal@unl.edu

## I. INTRODUCTION

Antiferromagnets offer a route to spintronic devices that operate with little stray magnetic fields and remain comparatively insensitive to external magnetic fields. Their fast spin dynamics also creates opportunities for switching and signal processing beyond the characteristic speeds of many ferromagnetic devices [1,2]. These benefits come with a practical difficulty: the nearly compensated magnetic moments of an antiferromagnet often make its order hard to detect electrically and hard to manipulate with conventional magnetic fields. Finding antiferromagnets that combine robust order with efficient electrical readout and control has therefore become a central materials challenge.

Altermagnets offer a way to address part of this challenge. Their oppositely oriented magnetic moments produce little or no net magnetization, while exchange interactions split the electronic bands in a pattern that alternates across momentum space [3–7]. This behavior arises because crystal rotations or mirror operations connect the opposite spin sublattices, rather than the translation or inversion-related symmetries that preserve spin degeneracy in conventional antiferromagnets. The resulting spin splitting can enable spin-polarized transport, spin-current generation, magnetoresistive readout, and electric-field control depending on the material and device geometry [8–21]. Candidate altermagnets span several chemical families, including chalcogenides such as MnTe, pnictides such as CrSb, oxychalcogenides such as $KV_2Se_2O$, and rutile fluorides and oxides such as $MnF_2$ and $RuO_2$ [22,23]. For each candidate, confirming the proposed magnetic ground state is essential to establishing whether the material is altermagnetic.

Rutile $RuO_2$ occupies a distinctive place among these candidates. It has been one of the first materials proposed as an altermagnet [24]. Bulk $RuO_2$ crystallizes in the rutile structure with space group $P4_2/mnm$. In its antiferromagnetic configuration, the two Ru sites within the crystallographic unit cell carry antiparallel magnetic moments. These opposite spin sublattices are connected by a fourfold rotation symmetry about the c axis (combined with half a unit cell translation) leading to a *d*-wave altermagnetic spin splitting in the electronic band structure. Initial theoretical calculations predicted a spin splitting of more than 1 eV along the $\Gamma-M$ direction [24]. Concrete device proposals soon followed, including a crystal-symmetry-driven anomalous Hall response for all-electrical readout [6], spin-splitter torque [9], and tunneling magnetoresistance in antiferromagnetic tunnel junctions [10].

However, conflicting experimental evidence regarding the magnetic state of *bulk* $RuO_2$ has already been reported well before the concept of altermagnetism was introduced. Early susceptibility measurements described it as a paramagnetic metal [25–27], while neutron and resonant X-ray studies reported evidence for antiferromagnetism [28,29]. More recent measurements on bulk crystals, including muon spectroscopy, diffraction, and quantum oscillations, have challenged that interpretation [30–33]. *Thin films* of $RuO_2$ present a different picture: magnetic and transport signatures vary with film orientation, substrate, thickness, and growth conditions [20,34–39]. These differences raise a concrete materials question: which structural and chemical changes can generate Ru moments, and which merely produce local magnetic signals without establishing an altermagnetic $RuO_2$ phase?

Epitaxial growth changes the lattice constants and Ru-O bonding, potentially bringing $RuO_2$ close to a magnetic instability. Vacancies and substitutional dopants can alter carrier density but also disrupt the local coordination of Ru. Interfaces introduce additional charge redistribution and structural relaxation. Because these effects often occur together in a film, their separate roles cannot be inferred reliably from a magnetic or transport signal alone. A comparative treatment of bulk material, strained films, defects, and interfaces is needed to identify conditions that favor an ordered Ru sublattice.

Here we use first-principles calculations and spin-susceptibility analysis to assess how epitaxial strain, substrate orientation, hole doping, Ru vacancies, Cr and Mn substitution, and $RuO_2/TiO_2$ interfaces affect magnetism in $RuO_2$. We compare nonmagnetic, ferromagnetic, and compensated antiferromagnetic states; examine representative substrate-induced lattice distortions; and distinguish moments on Ru from those centered on dopants. Our calculations find a nonmagnetic ground state for unstrained bulk $RuO_2$ without an imposed Hubbard correction. Appropriate epitaxial strain can instead favor Ru magnetic order, with the outcome strongly dependent on substrate and orientation. Hole doping and Cr or Mn substitution can strengthen the strain-induced state, whereas the local distortion caused by Ru vacancies can suppress it. Charge transfer at $RuO_2/TiO_2$ interfaces further modifies the layer-resolved Ru moments. These results identify strain as the principal control parameter in the systems studied and clarify why nominally similar $RuO_2$ samples can yield different magnetic signatures.

## II. METHODS

First-principles DFT calculations were performed using the Vienna *ab initio* Simulation Package (VASP) [40]. Ion-electron interactions were treated using the projector augmented wave method (PAW) [41,42]. The Perdew-Burke-Ernzerhof (PBE) functional was used to treat the exchange-correlation energy [43]. A plane-wave cutoff energy was set to 500 eV. An $11 \times 11 \times 16$ k-point mesh was applied and suitably reduced when supercells were considered. Atomic positions were relaxed until the force on each atom was less than $10^{-4}$ eV/Å, and self-consistency of the electronic structure calculations was achieved with a tolerance of $10^{-8}$ eV. Structural relaxation of the (101)-oriented structure was performed using Quantum ESPRESSO within the PAW method and PBE exchange–correlation functional, employing plane-wave and charge-density energy cutoffs of 80 and 640 Ry, respectively [44]. The resulting lattice parameters are listed in Table 2. The lattice mismatch along a given axis was calculated as $(a_{\mathrm{RuO_2}} - a_{\mathrm{sub}})/a_{\mathrm{RuO_2}}$, where $a_{\mathrm{RuO_2}}$ and $a_{\mathrm{sub}}$ denote the corresponding lattice parameters of $RuO_2$ and the substrate, respectively. The susceptibility calculations were performed using a Wannier tight-binding Hamiltonian constructed from Ru-d orbitals, with Wannierization carried out using the Wannier90 code [45].

Several computational tools, including FindSpinGroup [46], and Pymatgen [47] were used for crystallographic analysis and electronic structure visualization.

## III. RESULTS

### A. Brief Overview of Existing Experimental Studies

The first experimental claim arose from polarized neutron diffraction (ND), which attributed weak intensity at the forbidden (100) reflection to an ordered Ru moment of approximately $0.05\ \mu_B$ at room temperature [28]. Resonant X-ray scattering supported this interpretation and placed the Néel vector along the tetragonal *c* axis, although the very weak signals underlying both measurements later became central to the debate [29]. Most experiments now argue against magnetic order in the bulk $RuO_2$. Muon spin rotation on single crystals detected no spontaneous internal field and bounded the static moment near $5 \times 10^{-4}\ \mu_B$ [30], while a combined muon and neutron study reproduced this bound and attributed the reported (100) intensity to multiple scattering [31]. The Mössbauer spectroscopy and inelastic-scattering lattice dynamics placed independent limits on moment formation and correlation strength [48]. Most decisively, de Haas-van Alphen oscillations matched the nonmagnetic band structure and were incompatible with the altermagnetic state [33]. Bulk transport and refined single-crystal diffraction point the same way [32,49], and calculations indicate that stoichiometric $RuO_2$ does not order [50].

Independent bulk-sensitive probes reinforce this conclusion. Spin-resolved angle-resolved photoemission spectroscopy (ARPES) on single crystals and films found no predicted altermagnetic splitting [51], while another study resolved spin-degenerate bulk bands and Dirac-nodal-line surface states [52]. Infrared spectroscopy and nuclear quadrupole resonance likewise found no evidence for robust bulk order [53,54].

Experiments on epitaxially strained $RuO_2$ provided more signatures of magnetism than bulk crystals. Strained thin film on $TiO_2$ substrate demonstrated anomalous Hall response, which weakens upon strain relaxation [37]. A later study reported a transition near 500 K and an altermagnetic polar metallic phase in approximately 2-nm-thick strained films [34]. However, the absence of comparable signatures in thicker films highlights the importance of thickness, orientation, and strain [35].

Film spectroscopy is more mixed. Magnetic circular dichroism in ARPES from $RuO_2(110)/TiO_2(110)$ was interpreted as time-reversal symmetry breaking [55], whereas spin-resolved ARPES on another (110) film found no altermagnetic splitting [51]. In 2.7-nm-thick strained films, spin-resolved ARPES detected spin textures that the authors attributed to a magnetic state [56].

X-Ray Magnetic Linear Dichroism (XMLD) studies also disagree. Temperature-dependent Ru $M_3$-edge dichroism was attributed to Néel order along the c axis in epitaxial films [57], and a single-variant $RuO_2$ (101) film showed XMLD and spin-splitting magnetoresistance [58]. Conversely, X-ray linear dichroism in $TiO_2$-supported films showed little dependence on temperature or cooling field and was attributed to crystal-field anisotropy [59].

Spin-splitter torques, inverse spin-splitting signals, magnetoresistance, and Hall effects have been reported in films [20, 60–66]. Yet other studies of $TiO_2$-supported (100), and (101) films found no transport spin-splitting effect [67]. Terahertz measurements found either no inverse spin-splitting contribution [35] or a dominant inverse spin Hall signal [68]. These results require careful separation of magnetic, spin-orbit, and interfacial contributions.

Although calculations predict that hole doping can stabilize altermagnetism [50, 69], experiments on Cr-doped bulk $RuO_2$ found no long-range altermagnetic order. Instead, the holes remain localized near Cr and promote magnetic $CrO_2$ clusters [70]. Interfaces may be another possibility of supporting magnetism. Low-energy muon measurements confined static magnetism to within approximately 10 nm of the surface [71], while calculations predict that surface and interface symmetry breaking can generate Ru moments in the first few layers and produce two-dimensional altermagnetism in $RuO_2/TiO_2$ superlattices [72,73]. Thus, strain, interfacial reconstruction, and carrier redistribution likely act together in ultrathin $RuO_2$.

Table 1 summarizes the experimental evidence for and against magnetic order in bulk $RuO_2$ and films grown on different substrates, grouped by film orientation. While most bulk measurements favor a nonmagnetic ground state, the film results are mixed even for nominally similar orientations and substrates, highlighting the sensitivity of the reported signatures to sample conditions.

Most experimental studies of magnetism in $RuO_2$ have focused on bulk samples and thin films grown on $TiO_2$ and $Al_2O_3$ substrates. Other substrates include $MgF_2$, $SrTiO_3$, yttria-stabilized zirconia (YSZ), and MgO. Films grown on $TiO_2$ and $MgF_2$ typically exhibit a single structural domain, whereas those grown on $Al_2O_3$, $SrTiO_3$, and YSZ contain multiple structural domains [61, 62, 74]. This domain structure can affect some measurements such as the altermagnetic spin-splitting effect (ASSE). For example, the net ASSE vanishes in $RuO_2(100)$ films on $Al_2O_3$ when the three domains related by $C_3$ symmetry are equally populated, whereas a finite ASSE can occur in single-domain films on $TiO_2$ [75, 61].

We also investigated epitaxial strain conditions for substrates that have not previously been explored experimentally for altermagnetism in $RuO_2$ films. These include the rutile substrates $ZnF_2$ and $SnO_2$, which have been used for the growth of rutile compounds such as $MnF_2$, $FeF_2$ [76, 77, 78]. Specifically, the $RuO_2/SnO_2$ epitaxy has also been demonstrated at the nanoscale [79]. In addition, we considered forsterite, topaz, and alexandrite, which have been proposed as substrates for rutile materials [80]. Among these, forsterite $Mg_2SiO_4(010)$ provides reasonable epitaxial strains of 5.8% and −3.7% along the $RuO_2$ [010] and [001] directions, respectively, for a $RuO_2(100)$ film.

**Table 1:** Experimental reports on magnetic order in bulk and thin-film $RuO_2$. Film results are grouped by orientation and substrate and classified according to whether they support a nonmagnetic or paramagnetic state, or an altermagnetic interpretation. The abbreviations used in the table are as follows: neutron diffraction (ND), muon spin relaxation/rotation ($\mu$SR), angle-resolved photoemission spectroscopy (ARPES), anomalous Hall effect (AHE), planar Hall effect (PHE), magnetoresistance (MR), altermagnetic spin-splitting effect (ASSE), inverse altermagnetic spin-splitting effect (IASSE), inverse spin Hall effect (ISHE), quantum oscillatory spectroscopy (QOS), spin-splitting torque (SST), X-ray linear dichroism (XLD), X-ray magnetic linear dichroism (XMLD), photoemission electron microscopy (PEEM), spin-torque ferromagnetic resonance (ST-FMR), second-harmonic generation (SHG), magneto-optical (MO), and magneto-optical Kerr effect (MOKE).

| Film orientation /Substrate | Experiments supporting the nonmagnetic or paramagnetic order | Experiments supporting the altermagnetic order |
|---|---|---|
| Bulk $RuO_2$ | - μSR on single crystal ($m \approx 4.8 \times 10^{-4}\, \mu_B$) [30]<br>- μSR and ND study ($m \approx 1.14 \times 10^{-4}\, \mu_B$) [31]<br>- Polarized ND on single crystal ($m < 0.01\mu_B$) [32]<br>- Spin-resolved ARPES (spin-degenerate bands and Dirac nodal line surface states) [51,52]<br>- MR and AHE on single crystal (nonmagnetic semimetal) [81]<br>- Symmetry-sensitive torque magnetometry and magnetization measurements on single crystal [82]<br>- MR and AHE on single crystal (not strongly conclusive) [49]<br>- Infrared spectroscopy [53]<br>- DC susceptibility on single crystal [83] | - Polarized ND on anisotropic crystal ($m = 0.05\mu_B$ at room temp.) [28]<br>- Resonant X-ray scattering ($T_N > 300\ K$) [29]<br>- Soft X-ray ARPES (not conclusive) [85] |
| | - Specific heat and magnetic susceptibility [84]<br>- Magnetic QOS on single crystal [33]<br>- SHG and MO effects [34]<br>- Nuclear quadrupole resonance [54]<br>- Mössbauer spectroscopy, inelastic X-ray and neutron scattering [48] | |
| $RuO_2(001)$/<br>$TiO_2(001)$ | - | - 3rd order nonlinear Hall effect [60]<br>- XMLD on 15 nm epitaxial sample [57] |
| $RuO_2(100)$/<br>$TiO_2(100)$ | - Angle-dependent MR in Py/$RuO_2$ (not conclusive) [86]<br>- No IASSE; YIG spin Seebeck conversion is explained by anisotropic ISHE. [67]<br>- No IASSE signature in THz spin-to-charge conversion in Py/$RuO_2$ [35]<br>- No IASSE signature in THz emission in $RuO_2$(10 nm)/CoFeB [68] | - 3rd order nonlinear Hall effect [60]<br>- Separation of ISHE and IASSE voltages from spin Seebeck injection [61]<br>- Reversal of IASSE upon Néel vector switching through spin Seebeck measurements [87]<br>- XMLD on 15 nm epitaxial sample [57]<br>- Py-based ST-FMR in 1.5–20 nm films: larger T-odd SHC in coherently strained 4 nm films, smaller in partially relaxed films >7 nm [88] |
| $RuO_2(100)$/<br>$MgF_2(100)$ | - No $RuO_2$ magnon mode in magneto-Raman measurements [89] | - XMLD on 15 nm epitaxial sample [57] |
| $RuO_2(100)$/<br>$Al_2O_3(0001)$ | - | - XMLD detects current-induced Néel switching; ST-FMR reveals tunable ASSE-induced torque in non-epitaxial Py/$RuO_2$/Pt/$Al_2O_3$ [36]<br>- Separation of ISHE and IASSE voltages from spin Seebeck injection [61]<br>- Harmonic Hall measurements reveal enhanced field-like torque attributed to ASSE [90]<br>- ASSE induced large spin Hall angle in ST-FMR and spin pumping measurements [91] |
| $RuO_2(100)$/<br>$SrTiO_3(100)$ | - | - Incommensurate spin density wave (period of 14-20 nm) evidenced by abnormal SST in ST-FMR measurements [62]<br>- Spin flop transition in $SrTiO_3(100)$/$RuO_2(100)$/ MgO(110)/ $RuO_2(100)$ tunnel junction [92]<br>- Resonant X-ray scattering [29] |
| $RuO_2(100)$/<br>YSZ(100) | - | - ST-FMR detects ASSE-induced torque in $RuO_2$/Py. [63, 93] |
| $RuO_2(110)$/<br>$TiO_2(110)$ | - μSR and ND on 11 nm thick epitaxial sample ($m \approx 7.5 \times 10^{-4}\ \mu_B$) [31]<br>- Spin-resolved ARPES find no splitting larger than 10 meV [51]<br>- Exchange bias measurement suggests interfacial spin pinning over intrinsic magnetic order [89]<br>- Depth-resolved low-energy μSR on 33 nm thick film (magnetic order in near surface region $<$ 10 nm) [71]<br>- Strain-induced changes in oxygen vacancies generate spin disorder (does not establish altermagnetism) [94]<br>- Spin-resolved ARPES+DFT shows surface ferrimagnetism in 15 nm strain-relaxed film [95]<br>- XLD and PEEM: dichroism attributed to crystal-field anisotropy [59] | - Broken time-reversal symmetry in MCD in ARPES on 34 nm thick epitaxial sample [55]<br>- SHG, MO, and electrical measurements (time-reversal symmetry breaking in film of $<$ 4 nm thickness. No such effects were observed in strain-relaxed film of the 17 nm thickness) [34, 96]<br>- Broken time-reversal symmetry in spin-resolved ARPES on 2.7 nm thick sample [56]<br>- XMLD on 15 nm epitaxial sample [57]<br>- AHE by field induced reorientation of Neel vector in 9 nm thick sample [97]<br>- PHE and AHE on 40 nm thick sample [98]<br>- AHE in less than 4 nm films [37]<br>- Resonant elastic X-ray scattering in 21 nm thick film: (100) reflections are interpreted as signature of altermagnetism [100] |
| $RuO_2(110)$/<br>$MgF_2(110)$ | - Exchange bias measurement suggests interfacial spin disorder over intrinsic magnetic order [89] | - MR, anisotropic MR, and spin Hall MR in 30.6 nm thick film under strong magnetic field [99]<br>- XMLD on 15 nm epitaxial sample [57] |
| $RuO_2(110)$/<br>MgO(100) | - Strain-induced changes in oxygen vacancies generate spin disorder (does not establish altermagnetism) [94] | - XMLD, Hall resistance and ST-FMR measurements of Néel vector switching in Py/$RuO_2$/Pt/MgO sample (non-epitaxial) [36] |
| $RuO_2(101)$/<br>$TiO_2(101)$ | - YIG spin Seebeck conversion is explained by anisotropic ISHE, without requiring IASSE [67]<br>- No IASSE signature in THz spin-to-charge conversion in $Py/RuO_2$ [35] | - X-ray scattering in 18.6 nm thick film: (100) reflections are interpreted as signature of altermagnetism [100]<br>- Spin-split MR in $RuO_2$(3 nm)/Co [64]<br>- ST-FMR, damping modulation, and spin pumping reveal anisotropic ASSE and IASSE in epitaxial 15 nm thick $RuO_2$/Py [65] |
| $RuO_2(101)$/<br>$Al_2O_3(1\bar{1}02)$ | - Depth-resolved low-energy µSR on 30 nm thick sample (magnetic order in near surface region < 10 nm) [71] | - ST-FMR detects ASSE-induced torque in $RuO_2$/Py and field-free switching in $RuO_2$/Ru/Co/Pt [20]<br>- XMLD supports AFM order in single-variant film; MR in CoFeB/$RuO_2$ is attributed to ASSE [58]<br>- Spin Seebeck effect driven detection of spin-to-charge conversion attributed to IASSE [101]<br>- ASSE-attributed z-polarized torque enables robust field-free switching in $RuO_2$(16 nm)/Pt/Co/Pt [66] |

### B. Effects of Strain

The rutile phase of $RuO_2$ belongs to the tetragonal space group $P4_2/mnm$. In the conventional unit cell, Ru atoms are situated at the corner and body-center positions, whereas the O sublattice is specified by the fractional coordinates $\pm(u, u, 0)$ and $\pm(u + 1/2, -u + 1/2, 1/2)$, with $u$ denoting the internal parameter. Geometry optimization performed within PBE gives $a$ = 4.519 Å, $c$ = 3.125 Å, and $u$ = 0.3057, while HSE06 yields $a$ = 4.463 Å, $c$ = 3.086 Å, and $u$ = 0.3030. These results are comparable with the corresponding experimental values of $a$ = 4.492 Å, $c$ = 3.107 Å, and $u$ = 0.3058 [102,103]. For bulk calculations, we use experimental structural parameters and PBE functional, unless stated otherwise.

The DFT calculation for the bulk $RuO_2$ structure reveals the nonmagnetic ground state. An altermagnetic solution emerges within DFT+U only when the effective Hubbard interaction, $U_{\mathrm{eff}} = U - J$, exceeds 1 eV, as shown in Fig. S1(a) of the Supplemental Material [104]. At $U_{\mathrm{eff}} = 1.25$ eV, the calculated magnetic moment is approximately 0.7 $\mu_B$ per Ru atom. Such a value of $U_{\mathrm{eff}}$ is relatively large for $RuO_2$, which is generally regarded as an itinerant, weakly correlated 4d-electron metal [48,50,53,69,105]. We therefore conclude that bulk $RuO_2$ has a nonmagnetic ground state, consistent with the majority of experimental studies summarized in Table 1. More physical insights from the Stoner model are given in section III C.

Since the DFT tends to overestimate magnetic ordering in itinerant electron systems [106,107], we do not introduce a Hubbard correction when determining the ground state in this work. Correlation effects may still be relevant in itinerant materials and can, in principle, be treated using a small Hubbard interaction within DFT+U or more advanced approaches such as dynamical mean-field theory (DMFT) [108]. However, for $RuO_2$ the nature of the ground state remains controversial, and an appropriate value of the Hubbard interaction has not been firmly established. Moreover, because the inclusion of Hubbard parameters generally favors moment formation, a state that is magnetic at the DFT level would most likely remain magnetic, and may exhibit an enhanced magnetic moment within DFT+U. We therefore use conventional DFT, without an empirical Hubbard correction, to assess the intrinsic tendency of $RuO_2$ toward magnetic ordering.

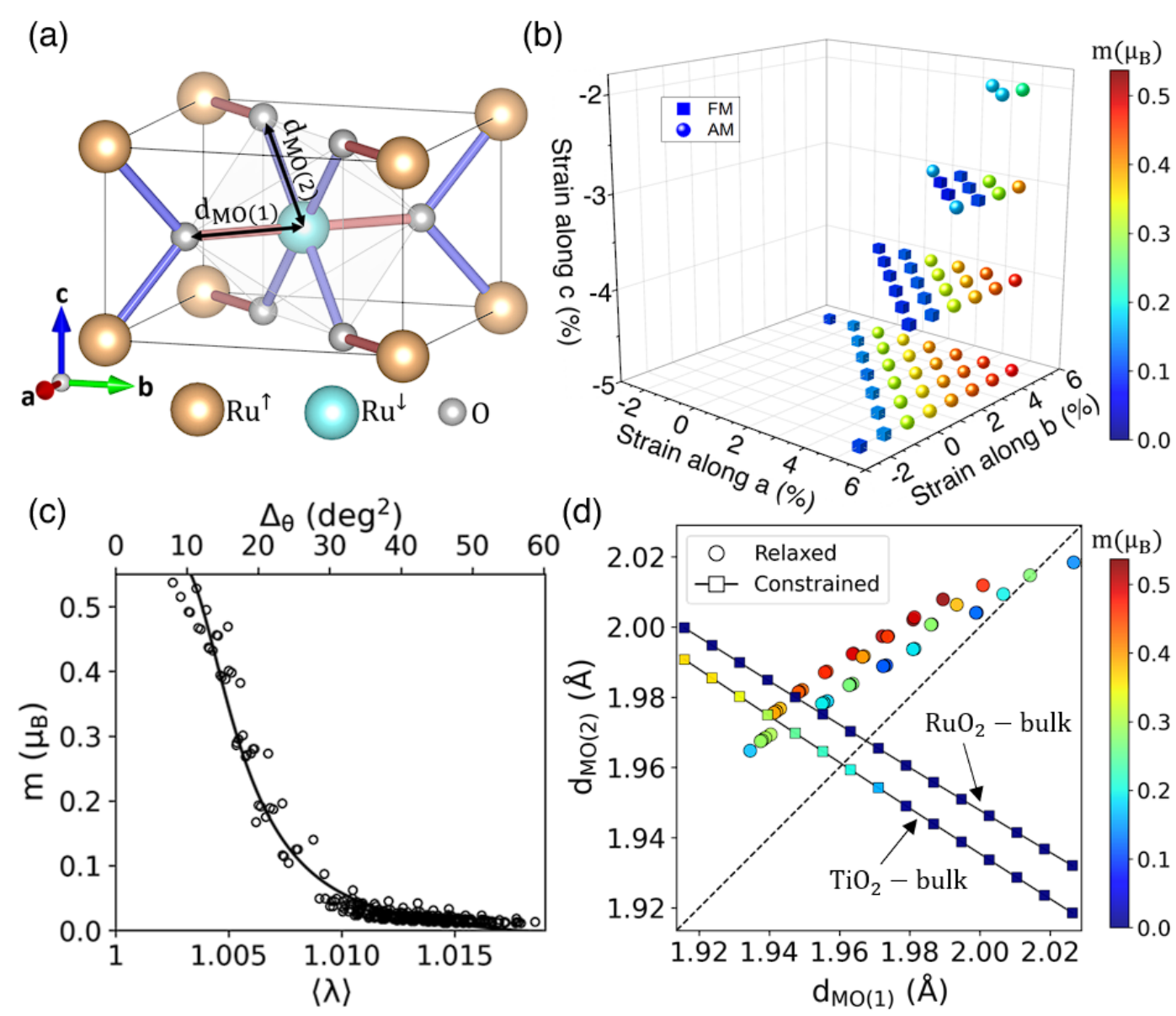

**Figure 1:** (a) Crystallographic unit cell of rutile $RuO_2$. Blue and red Ru-O bonds distinguish the two inequivalent $RuO_6$ octahedral bonds. (b) The Ru magnetic moments under strain along the crystallographic axes. Spheres and cubes indicate altermagnetic and ferromagnetic ground states, respectively. (c) The Ru magnetic moments for all structures as functions of the mean quadratic elongation $\langle\lambda\rangle$ and the octahedral cis-angle variance $\Delta_\theta$. The solid line is a guide to the eye. (d) Magnetic moment distribution in the $(d_{\mathrm{MO}(1)}, d_{\mathrm{MO}(2)})$ plane. Scatter points represent all strained structures considered in (b). Solid lines correspond to constrained variations of the internal parameter u, which changes $d_{\mathrm{MO}(1)}$ and $d_{\mathrm{MO}(2)}$ while keeping the lattice parameters fixed at the bulk values of $RuO_2$ or $TiO_2$.

**Table 2:** Structural parameters, local Ru magnetic moments, and energy differences between the altermagnetic and nonmagnetic phases for $RuO_2$ films with different crystallographic orientations on various substrates.

| Film/substrate orientation | Epitaxial relationship | In-plane lattice parameters (lattice mismatch) | Relaxed out-of-plane lattice parameter (lattice mismatch) | $\langle\lambda\rangle$ | $\Delta_\theta(\mathrm{deg}^2)$ | $m(\mu_B)$ | $E_{AM} - E_{NM}$ (meV/f.u.) |
|---|---|---|---|---|---|---|---|
| $RuO_2$(001) /$TiO_2$(001) | [100]//[100] | 4.593 Å (2.25 %) | 3.044 Å (–2.03 %) | 1.0101 | 32.624 | 0.0 | 0.0 |
| | [010]//[010] | 4.593 Å (2.25 %) | | | | | |
| $RuO_2$(001) /$MgF_2$(001) | [100]//[100] | 4.628 Å (3.03 %) | 3.006 Å (–3.25 %) | 1.0069 | 22.517 | 0.149 | –0.047 |
| | [010]//[010] | 4.628 Å (3.03 %) | | | | | |
| $RuO_2$(001) /$SnO_2$(001) | [100]//[100] | 4.737 Å (5.45 %) | 2.859 Å (–7.98 %) | 1.0004 | 0.918 | 0.499 | –6.032 |
| | [010]//[010] | 4.737 Å (5.45 %) | | | | | |
| $RuO_2$(001) /$ZnF_2$(001) | [100]//[100] | 4.703 Å (4.70 %) | 2.906 Å (–6.47 %) | 1.0014 | 4.399 | 0.523 | –19.046 |
| | [010]//[010] | 4.703 Å (4.70 %) | | | | | |
| $RuO_2$(100) /$TiO_2$(100) | [010]//[010] | 4.593 Å (2.25 %) | 4.736 Å (5.43 %) | 1.0039 | 12.679 | 0.475 | –11.236 |
| | [001]//[001] | 2.959 Å (–4.76 %) | | | | | |
| $RuO_2$(100) /$MgF_2$(100) | [010]//[010] | 4.625 Å (2.96 %) | 4.548 Å (1.25 %) | 1.0107 | 34.696 | 0.0 | 0.0 |
| | [001]//[001] | 3.052 Å (–1.77 %) | | | | | |
| $RuO_2$(100)/ $Al_2O_3$(0001) | $[010]//[11\bar{2}0]$ | 4.750 Å (5.74 %) | 4.899 Å (9.06 %) | 1.0012 | 3.703 | 0.0 | 0.0 |
| | $[001]//[1\bar{1}00]$ | 2.740 Å (–11.81 %) | | | | | |
| $RuO_2$(100) /$Mg_2SiO_4$(010) | [010]//[100] | 4.753 Å (5.81 %) | 4.524 Å (0.71 %) | 1.0063 | 20.300 | 0.254 | –0.587 |
| | [001]//[001] | 2.991 Å (–3.73 %) | | | | | |
| $RuO_2$(100) /$SrTiO_3$(100) | [011]//[011] | 4.573 Å (1.80 %) | 4.541 Å (1.09 %) | 1.0136 | 43.859 | 0.0 | 0.0 |
| | | 3.084 Å (–0.74 %) | | | | | |
| $RuO_2$(110) /$TiO_2$(110) | $[1\bar{1}0]//[1\bar{1}0]$ | 6.495 Å (2.24 %) | 6.678 Å (5.12 %) | 1.0032 | 9.111 | 0.483 | –10.899 (FiM) |
| | [001]//[001] | 2.959 Å (–4.76 %) | | 1.0054 | 17.702 | 0.364 | |
| $RuO_2$(110) /$MgF_2$(110) | $[1\bar{1}0]//[1\bar{1}0]$ | 6.541 Å (2.96 %) | 6.435 Å (1.30 %) | 1.0117 | 38.215 | 0.005 | 0.001 (FiM) |
| | [001]//[001] | 3.052 Å (–1.77 %) | | 1.0097 | 30.939 | 0.006 | |
| $RuO_2$(101) /$TiO_2$(101) | $[10\bar{1}]//[10\bar{1}]$ | 5.464 Å (0.04 %) | 5.420 Å (–0.76 %) | 1.0160 | 51.247 | 0.006 | 0.001 |
| | [010]//[010] | 4.593 Å (2.24 %) | | | | | |
| $RuO_2$(101) /$MgF_2$(101) | $[10\bar{1}]//[10\bar{1}]$ | 5.541 Å (1.45 %) | 5.399 Å (–1.15 %) | 1.0148 | 47.530 | 0.0 | 0.0 |
| | [010]//[010] | 4.625 Å (2.96 %) | | | | | |
| $RuO_2$(101) /$SnO_2$(101) | $[10\bar{1}]//[10\bar{1}]$ | 5.709 Å (4.52 %) | 5.338 Å (–2.27 %) | 1.0124 | 40.277 | 0.0 | 0.0 |
| | [010]//[010] | 4.737 Å (5.45 %) | | | | | |
| $RuO_2$(101)/ $Al_2O_3$($1\bar{1}02$) | $[\bar{1}01]//[\bar{1}101]$ | 5.123 Å (–6.20 %) | 5.327 Å (–2.47 %) | 1.0215 | 66.809 | 0.0 | 0.0 |
| | $[010]//[11\bar{2}0]$ | 4.750 Å (5.74 %) | | | | | |

To systematically investigate strain effects, we apply strains ranging from –5% to +5% along each of the three crystallographic axes, referenced to the experimental bulk lattice parameters. For each strain case, the internal atomic coordinates are fully relaxed. We find that magnetism appears only under *c*-axis compression and tensile strain along the *a* and *b* axes. Figure 1(b) shows the magnetic moment of the energetically preferred state among the ferromagnetic and altermagnetic configurations. A weakly magnetic ferromagnetic phase occurs within a narrow strain region near the transition from the nonmagnetic to the altermagnetic state. Within the investigated strain range, increasing the compressive strain along *c*, or the tensile strain along *a* and *b*, enhances the magnetic moment and thus stabilizes the altermagnetic state.

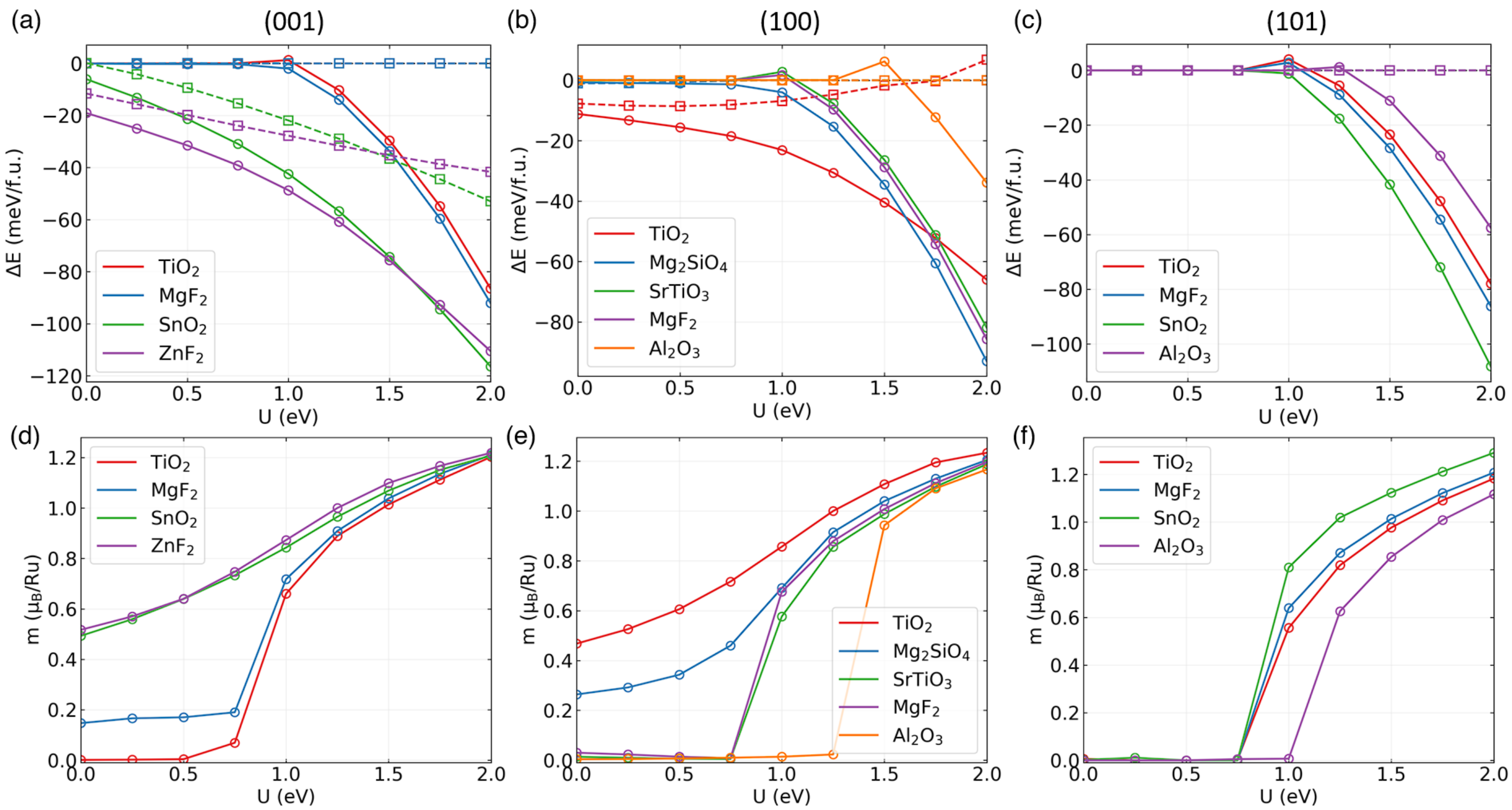

**Figure 2:** (a-c) Energy differences relative to the nonmagnetic phase for the altermagnetic (solid lines) and ferromagnetic (dashed lines) phases as functions of Hubbard U for the (001), (100), and (101) orientations, respectively. (d-f) The Ru magnetic moments in the altermagnetic phase as functions of Hubbard U for the corresponding orientations.

The two inequivalent metal-oxygen bond lengths, $d_{\mathrm{MO}(1)}$ and $d_{\mathrm{MO}(2)}$, defined in Fig. 1(a), provide additional insight into the structural factors. Bonds in bulk $RuO_2$ relax to $d_{\mathrm{MO}(1)} < d_{\mathrm{MO}(2)}$, whereas opposite relation, $d_{\mathrm{MO}(1)} > d_{\mathrm{MO}(2)}$, occurs in bulk $TiO_2$ [109]. In Fig. 1(d), the scattered data points represent the magnetic moments calculated for all strained structures after atomic relaxation, plotted in the $(d_{\mathrm{MO}(1)}, d_{\mathrm{MO}(2)})$ plane. The relaxation maintains $d_{\mathrm{MO}(1)} < d_{\mathrm{MO}(2)}$ in most cases, similar to the unstrained bulk $RuO_2$. The solid curve shows the effect of varying the internal parameter at fixed $TiO_2$ lattice constants, revealing a suppression of the magnetic moment as $d_{\mathrm{MO}(1)}/d_{\mathrm{MO}(2)}$ increases. However, analogous variation of internal parameter at the bulk $RuO_2$ lattice constants shows that reducing $d_{\mathrm{MO}(1)}/d_{\mathrm{MO}(2)}$ is insufficient to generate a magnetic state. Therefore, although the metal-oxygen bond length ratio influences the magnitude of the magnetic moment, strain is the primary structural parameter controlling magnetism in $RuO_2$.

To further quantify the relationship between structural distortion and magnetism, we evaluate two distortion measures for the $RuO_6$ coordination octahedron: the mean quadratic elongation $\langle\lambda\rangle$ [110,111]

$$\langle\lambda\rangle = \frac{1}{6}\sum_{i=1}^{6}\left(\frac{d_i}{d_0}\right)^2,$$

and the variance of octahedral cis-angle $\Delta_\theta$

$$\Delta_\theta = \frac{1}{12}\sum_{i=1}^{12}(\theta_i - 90°)^2 .$$

Here $d_i$ denotes the Ru-O bond lengths of $RuO_6$, $d_0$ denotes the Ru-O bond length of ideal $RuO_6$ octahedron of the same volume, and $\theta_i$ denotes the O-Ru-O octahedron cis angle. Fig. 1(c) shows that magnetic states appear for $\langle\lambda\rangle$ between approximately 1.002 and 1.009, and the magnetic moment tends to increase as $\langle\lambda\rangle$ decreases. A similar trend is found for $\Delta_\theta$ over the range 8 deg$^2$ to 30 deg$^2$. Thus, within the strain regime explored here, a larger magnetic moment correlates with reduced octahedral distortion. Further reductions in $\langle\lambda\rangle$ and $\Delta_\theta$ would require unrealistically large strains. Moreover, this correlation should not be extrapolated beyond the investigated structural range, as demonstrated in Table 2 for the specific case of $RuO_2$ (100) grown on $Al_2O_3$ (0001). This correlation between the octahedral-distortion measures and the magnetic moment will be useful for tracking the emergence and suppression of magnetism caused by structural distortions in the doping and interface calculations discussed in later sections.

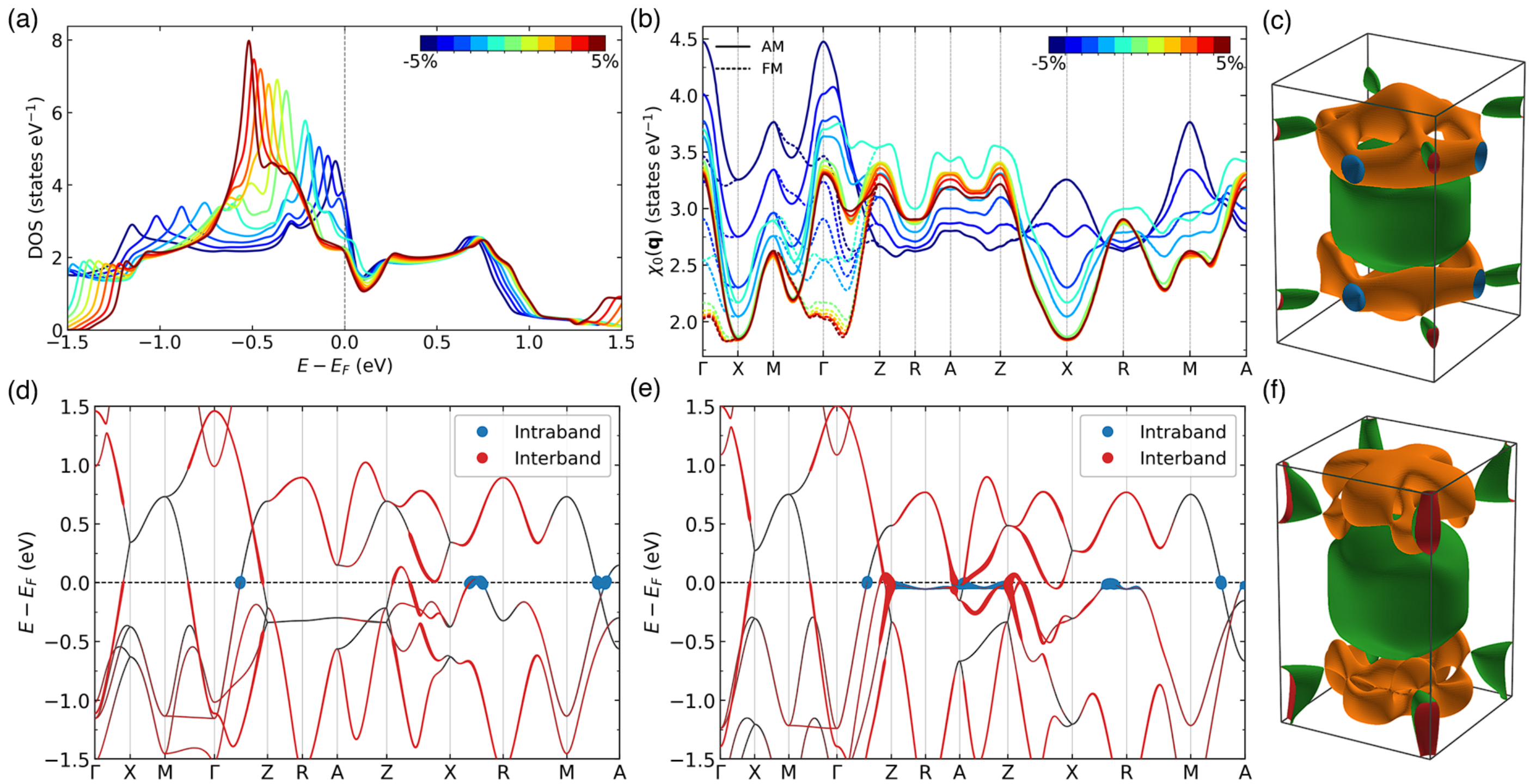

**Figure 3:** Electronic structure and magnetic susceptibility of $RuO_2$ under 5% tensile strain along both the *a* and *b* axes, with *c*-axis strain varied from −5% to 5%. (a) Nonmagnetic density of states per spin per unit cell. Colors indicate the *c*-axis strain. (b) Altermagnetic (solid lines) and ferromagnetic (dashed lines) susceptibilities along a high-symmetry path in the Brillouin zone. (c, f) Nonmagnetic Fermi surfaces for *c*-axis strains of 0% and −5%, respectively. (d, e) Band-resolved intraband and interband contributions to $\chi_0^{\mathrm{AM}}(\Gamma)$ for *c*-axis strains of 0% and −5%, respectively.

We next examine epitaxially strained $RuO_2$ films with different orientations. For each film, we fix the in-plane lattice parameters to those of the corresponding substrate and relax the out-of-plane lattice parameter and internal atomic coordinates. For $RuO_2$ (101), we also relax the interaxial angle. The substrate parameters and the resulting structural and magnetic properties are summarized in Table 2. The altermagnetic band structures of the strained systems that exhibit an altermagnetic ground state without a Hubbard $U$ correction (Supplemental Material, Sec. III [104]).

For the (001) orientation, our DFT calculations predict an altermagnetic ground state in $RuO_2$ on the previously unexplored substrates $MgF_2$, $SnO_2$, and $ZnF_2$. The magnetic moments are approximately 0.5 $\mu_B$ for $SnO_2$, and $ZnF_2$, and 0.15 $\mu_B$ for $MgF_2$. As shown in Figs. 2(a, d), including Hubbard $U$ in the calculation further stabilizes the altermagnetic phase, making the energy difference between the altermagnetic and nonmagnetic states, $E_{\mathrm{AM}} - E_{\mathrm{NM}}$, more negative, and enhances the magnetic moments. In contrast, $RuO_2$ on the experimentally investigated $TiO_2$ substrate has a nonmagnetic ground state, with altermagnetism emerging only for $U > 1$ eV, similar to bulk $RuO_2$. These results suggest that $SnO_2$, and $ZnF_2$ are more promising substrates than $TiO_2$ for stabilizing altermagnetism in $RuO_2$ (001) film, with $ZnF_2$ particularly favorable for the reason discussed in the next section.

The (100) orientation of $RuO_2$ has been investigated extensively in experiments, primarily using $TiO_2$ and α-$Al_2O_3$ substrates. As shown in Figs. 2(b, e), our calculations predict an altermagnetic ground state for $RuO_2$(100)/$TiO_2$(100) with a magnetic moment of 0.475 $\mu_B$, and increasing to 0.866 $\mu_B$ at $U$ = 1eV. Although several experiments have been interpreted as evidence for altermagnetism in $RuO_2$, others report no magnetic order, and some remain inconclusive. In contrast, all experimental studies of $RuO_2$(100)/$Al_2O_3$(0001) report an altermagnetic ground state, whereas our calculations predict a nonmagnetic ground state, with the altermagnetic phase becoming stable only for $U > 1.5$ eV. This heterostructure has a very large lattice mismatch, corresponding to approximately 11.8% compressive strain along the $RuO_2$ [001] direction. Such a large mismatch is unlikely to persist throughout the film and may be relieved by misfit dislocations and spatially nonuniform relaxation or may lead to non-epitaxial growth [36]. Experimental studies have reported strain relaxation and spatially inhomogeneous strain in $RuO_2$ (100) films grown on $Al_2O_3$ (0001), with contributions from domain boundaries and growth morphology [112, 113]. Such local strain variations could create regions where magnetic order is favored, potentially explaining the difference between calculations for ideal, coherently strained films and experimental observations.

The $RuO_2$(100)/$SrTiO_3$(100) has also been investigated experimentally, with reports supporting altermagnetism and incommensurate spin-density wave order. Our DFT calculations instead favor a nonmagnetic ground state among the configurations considered. The possibility of an interaction driven spin-density wave instability is discussed in the next section. For the $RuO_2$(100)/$MgF_2$(100), our calculations predict a nonmagnetic ground state, with altermagnetism stabilized only for $U > 1$ eV. In contrast, for $RuO_2$(100)/$Mg_2SiO_4$(010), we find an altermagnetic ground state with a magnetic moment of $0.25\mu_B$ per Ru atom, which increases significantly upon inclusion of Hubbard $U$. These results motivate experimental investigations of $RuO_2$(100)/$Mg_2SiO_4$(010) heterostructures.

For $RuO_2$ (110) orientation, the epitaxial strain breaks the symmetry that relates the two opposite-spin Ru sublattices. Consequently, the sublattice moments become unequal, resulting in ferrimagnetic ground state with a finite net moment. Calculation predicts the magnetic moments of $0.48\,\mu_B$ and $-0.36\,\mu_B$ in the case of $RuO_2$(110)/$TiO_2$(110). The effect of Hubbard $U$ and doping is shown in Sec II of Supplemental Material [104]. Experimental findings for this orientation are likewise controversial.

The (101) orientation of $RuO_2$ also has been extensively studied on $TiO_2$ and $\alpha$-$Al_2O_3$ substrates, although the magnetic ground state remains controversial in both cases. For this orientation, our calculations predict a nonmagnetic ground state for all substrates considered, with the altermagnetic phase stabilized only for Hubbard $U > 1$ eV [Figs. 2(c, f)].

### C. Magnetic Susceptibility

To gain further physical insight into strain-induced magnetism in $RuO_2$, we use the Stoner framework to analyze the spin susceptibility and assess the instability of the nonmagnetic state toward magnetic ordering. Although magnetic susceptibility of $RuO_2$ has been calculated previously [114,115], the effects of strain on the competing ferromagnetic and altermagnetic instabilities have not been systematically studied. We therefore calculate the bare generalized susceptibilities for the ferromagnetic and altermagnetic states using a tight-binding model constructed from Wannier functions spanning the Ru $d$ orbitals.

We introduce the composite index $\mu = (l, \alpha)$, where $l$ denotes the orbital and $\alpha$ denotes sublattice, giving total $N_0$

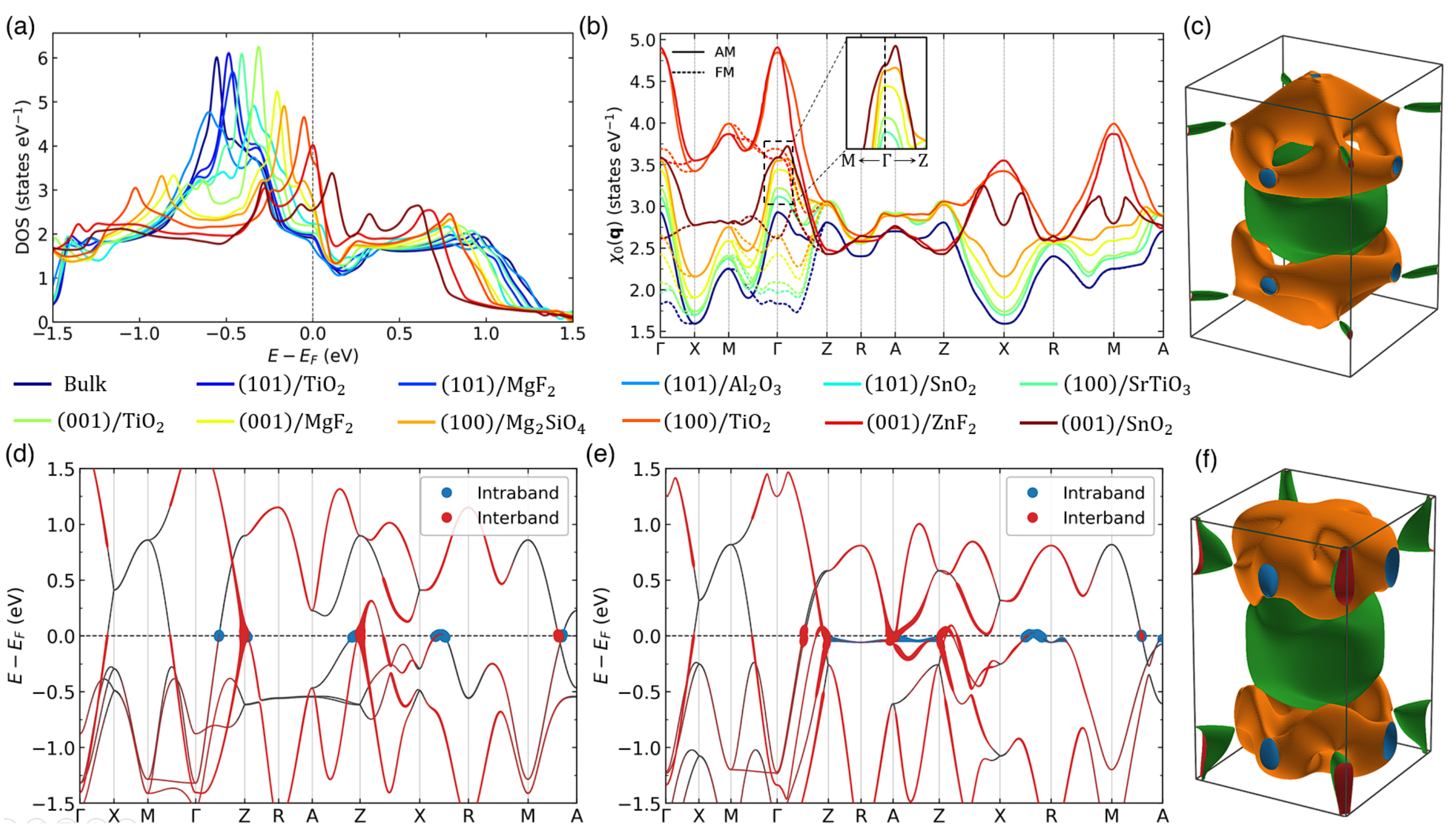

**Figure 4:** Electronic structure and magnetic susceptibility of $RuO_2$ constrained to different substrates and different orientations. (a) Nonmagnetic density of states per spin per unit cell. (b) Altermagnetic (solid lines) and ferromagnetic (dashed lines) susceptibilities along a high-symmetry path in the Brillouin zone. (c, f) Nonmagnetic Fermi surfaces for bulk $RuO_2$ (c) and $RuO_2$(100)/$TiO_2$(100) (f), respectively. (d, e) Band-resolved intraband and interband contributions to $\chi_0^{\mathrm{AM}}(\Gamma)$ for bulk $RuO_2$ (d) and $RuO_2$(100)/$TiO_2$(100) (e), respectively.

degrees of freedom. The non-interacting one-body Hamiltonian is $H_0 = \sum_{\boldsymbol{k}} \Psi_{\boldsymbol{k}}^\dagger H(\boldsymbol{k}) \Psi_{\boldsymbol{k}}$, where $\Psi_{\boldsymbol{k}} = \left(c_{\boldsymbol{k}1}, c_{\boldsymbol{k}2}, \dots, c_{\boldsymbol{k}N_0}\right)^T$, and $c_{\boldsymbol{k}\mu}$ is the fermionic annihilation operator associated with the state $\mu$ at momentum $\boldsymbol{k}$. State-dependent bare susceptibility is defined as

$$\chi_0^{\tau_M}(\boldsymbol{q}, i\Omega_n) = -\int_0^\beta d\tau\, e^{-i\Omega_n \tau} \left\langle T_\tau\, S_{\boldsymbol{q}}(\tau) S_{-\boldsymbol{q}}(0)\right\rangle,$$

where $S_{\boldsymbol{q}} = \frac{1}{N}\sum_{\boldsymbol{k}} \Psi_{\boldsymbol{k}}^\dagger \tau_M \Psi_{\boldsymbol{k}}$, $\tau_M = \tau_0$ for the ferromagnetic state, or $\tau_M = \tau_z$ for the altermagnetic state, $\beta$ is inverse temperature and $i\Omega_n$ are bosonic Matsubara frequencies [114, 116]. Using the Wick's theorem, the bare susceptibility can be written as [114,116,117]

$$\chi_0^{\tau_M}(\boldsymbol{q},\, i\Omega_n) = \sum_{\mu_1\mu_2\mu_3\mu_4} (\tau_M)_{\mu_1\mu_2} (\tau_M)^*_{\mu_3\mu_4} \left[\chi_0(\boldsymbol{q},\, i\Omega_n)\right]^{\mu_1\mu_2}_{\mu_3\mu_4},$$

where the bare susceptibility matrix is given by

$$\left[\chi_0(\boldsymbol{q},\, i\Omega_n)\right]^{\mu_1\mu_2}_{\mu_3\mu_4} = -\frac{1}{N\beta}\sum_{\boldsymbol{k}, i\omega_n} G^0_{\mu_1\mu_3}(\boldsymbol{k}+\boldsymbol{q}, i\omega_n + i\Omega_n) \times G^0_{\mu_2\mu_4}(\boldsymbol{k}, i\omega_n).$$

The bare Green's function takes the form

$$G^0_{\mu_1\mu_2}(\boldsymbol{k}, i\omega_n) = \sum_m \frac{a_m^{\mu_1}(\boldsymbol{k}) a_m^{\mu_2 *}(\boldsymbol{k})}{i\omega_n - E_m(\boldsymbol{k})},$$

where $a_m(\boldsymbol{k})$ denotes an eigenvector of $H(\boldsymbol{k})$, with components $a_m^\mu(\boldsymbol{k}) = \langle \mu | m\boldsymbol{k}\rangle$, and corresponding energy eigenvalues $E_m(\boldsymbol{k})$. Performing the Matsubara-frequency sum then yields the bare susceptibility matrix in terms of the Lindhard kernel,

$$\left[\chi_0(\boldsymbol{q}, i\Omega_n)\right]^{\mu_1\mu_2}_{\mu_3\mu_4} = \frac{1}{N}\sum_{\boldsymbol{k},m,n} a_n^{\mu_1 *}(\boldsymbol{k}) a_n^{\mu_3}(\boldsymbol{k}) a_m^{\mu_2}(\boldsymbol{k}+\boldsymbol{q}) a_m^{\mu_4 *}(\boldsymbol{k}+\boldsymbol{q}) \times \frac{f\left(E_m(\boldsymbol{k}+\boldsymbol{q})\right) - f\left(E_n(\boldsymbol{k})\right)}{i\Omega_n + E_n(\boldsymbol{k}) - E_m(\boldsymbol{k}+\boldsymbol{q})}.$$

Taking $i\Omega_n \to 0$, we obtain the static bare susceptibility

$$\chi_0^{\tau_M}(\boldsymbol{q}) = \frac{1}{N}\sum_{\boldsymbol{k},m,n} \left|\langle m, \boldsymbol{k}+\boldsymbol{q} | \tau_M | n, \boldsymbol{k}\rangle\right|^2 \times \frac{f\left(E_m(\boldsymbol{k}+\boldsymbol{q})\right) - f\left(E_n(\boldsymbol{k})\right)}{E_n(\boldsymbol{k}) - E_m(\boldsymbol{k}+\boldsymbol{q})}.$$

A pronounced bare-susceptibility peak at the Γ point in the altermagnetic state, characterized by $\chi_0^{\mathrm{AM}}(\Gamma) > \chi_0^{\mathrm{FM}}(\Gamma)$, indicates a possible altermagnetic instability. A dominant peak at the Γ point in the ferromagnetic state, characterized by $\chi_0^{\mathrm{AM}}(\Gamma) < \chi_0^{\mathrm{FM}}(\Gamma)$, signals a tendency toward ferromagnetic order. A dominant peak at a finite wave vector, away from the Γ point, instead suggests a possible spin-density-wave instability. The $\chi_0^{\mathrm{FM}}(\Gamma)$ contains only intraband contributions, whereas $\chi_0^{\mathrm{AM}}(\Gamma)$ includes both intraband and interband terms [114]. The intraband contribution arises only from states at the Fermi level. The interband contribution involves terms $\sim\left(E_m(\boldsymbol{k}) - E_n(\boldsymbol{k})\right)^{-1}$, where $E_m(\boldsymbol{k})$ and $E_n(\boldsymbol{k})$ denote unoccupied and occupied band energies, respectively. As a result, closely spaced bands on opposite sides of the Fermi level contribute strongly to $\chi_0^{\mathrm{AM}}(\Gamma)$.

We first examine structures under 5% in-plane tensile strain, varying the *c*-axis strain from +5% to −5%. This range covers the conditions under which the largest magnetic moments are obtained [Fig. 1(b)]. As the *c*-axis strain changes from tensile to compressive, the peak in the density of states shifts toward the Fermi level [Fig. 3(a)]. The corresponding susceptibilities satisfy $\chi_0^{\mathrm{AM}}(\Gamma) > \chi_0^{\mathrm{FM}}(\Gamma)$ for all strains considered [Fig. 3(b)]. For *c*-axis strains between −4% and −2%, the maximum of $\chi_0^{\mathrm{AM}}$ occurs at a finite wave vector along the Γ − Z path, suggesting a possible spin-density-wave instability. For remaining strain values, the peak lies at Γ and becomes more pronounced as the compressive *c*-axis strain increases, indicating an enhanced tendency toward altermagnetic order. As shown in Figs. 3 (d, e), the interband contributions to $\chi_0^{\mathrm{AM}}(\Gamma)$ are significantly larger at a *c*-axis strain of −5% than at 0%.

Next, we calculate the susceptibilities of $RuO_2$ under the epitaxial strain conditions imposed by the different substrates and orientations. In bulk $RuO_2$, the density-of-states peak lies approximately 0.6 eV below the Fermi level [Fig. 4(a)]. Peak shifts toward the Fermi level for all substrate-induced strain conditions considered, with the largest shifts occurring for $RuO_2$(100)/$TiO_2$(100), and $RuO_2$(001)/$ZnF_2$(001). As shown in Figs. 4(d, e), both the intraband and interband contributions to $\chi_0^{\mathrm{AM}}(\Gamma)$ are significantly enhanced in $RuO_2$(100)/$TiO_2$(100) compared with bulk $RuO_2$. The band-resolved contributions to $\chi_0^{\mathrm{AM}}(\Gamma)$ and the nonmagnetic Fermi surfaces for the remaining substrate-induced strain conditions are presented in Sec. IV of the Supplemental Material [104]. In every case, $\chi_0^{\mathrm{AM}}(\Gamma) > \chi_0^{\mathrm{FM}}(\Gamma)$ [Fig. 4(b)], with $RuO_2$(100)/$TiO_2$(100), and $RuO_2$(001)/$ZnF_2$(001) exhibiting the strongest peak in $\chi_0^{\mathrm{AM}}(\mathbf{q})$, both located at Γ, consistent with altermagnetic ground states discussed in the previous section. For $RuO_2$(100)/$Mg_2SiO_4$(010), and $RuO_2$(001)/$SnO_2$(001) the maxima of $\chi_0^{\mathrm{AM}}(\mathbf{q})$ occur away from Γ along the $\Gamma - \mathrm{Z}$ path, at $q_z = 0.1017\frac{2\pi}{c}$. This suggests a tendency toward spin-density-wave instability.

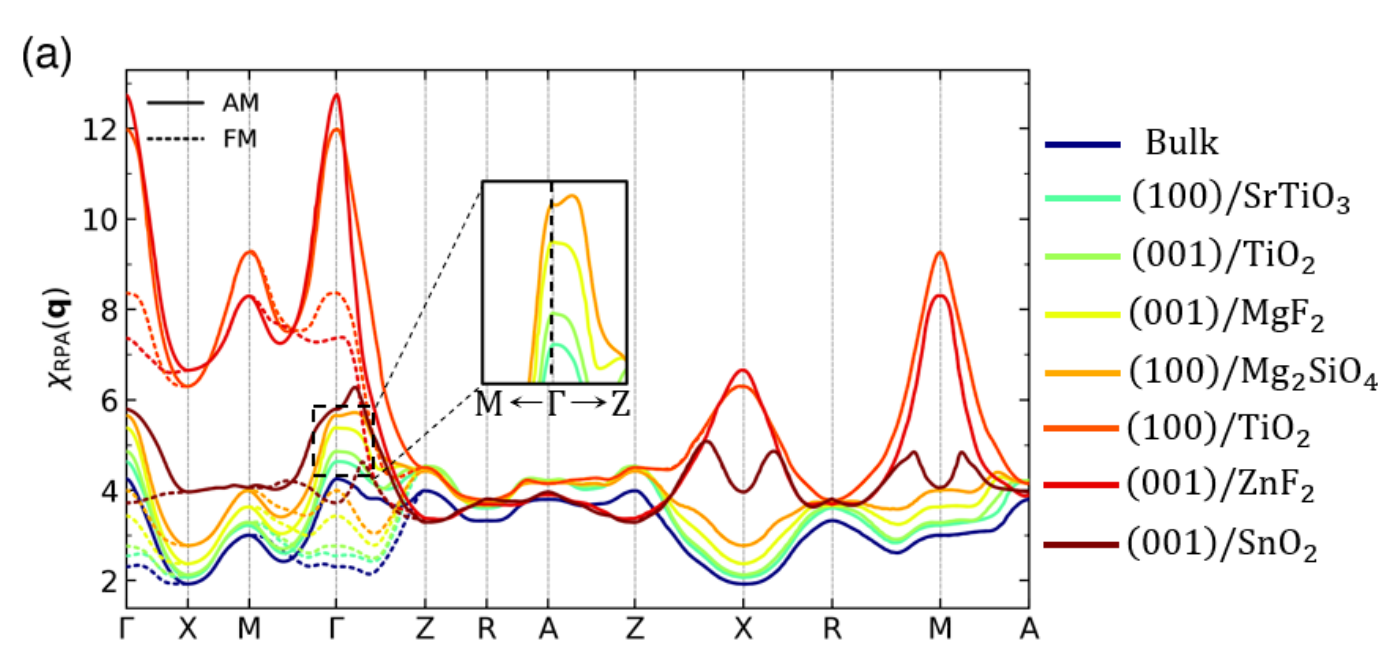

**Figure 5:** Altermagnetic (solid lines) and ferromagnetic (dashed lines) RPA susceptibilities with $U = 0.5\,eV$ and $J/U = 0.15$ for the $RuO_2$ under the epitaxial strain conditions imposed by the different substrates and orientations.

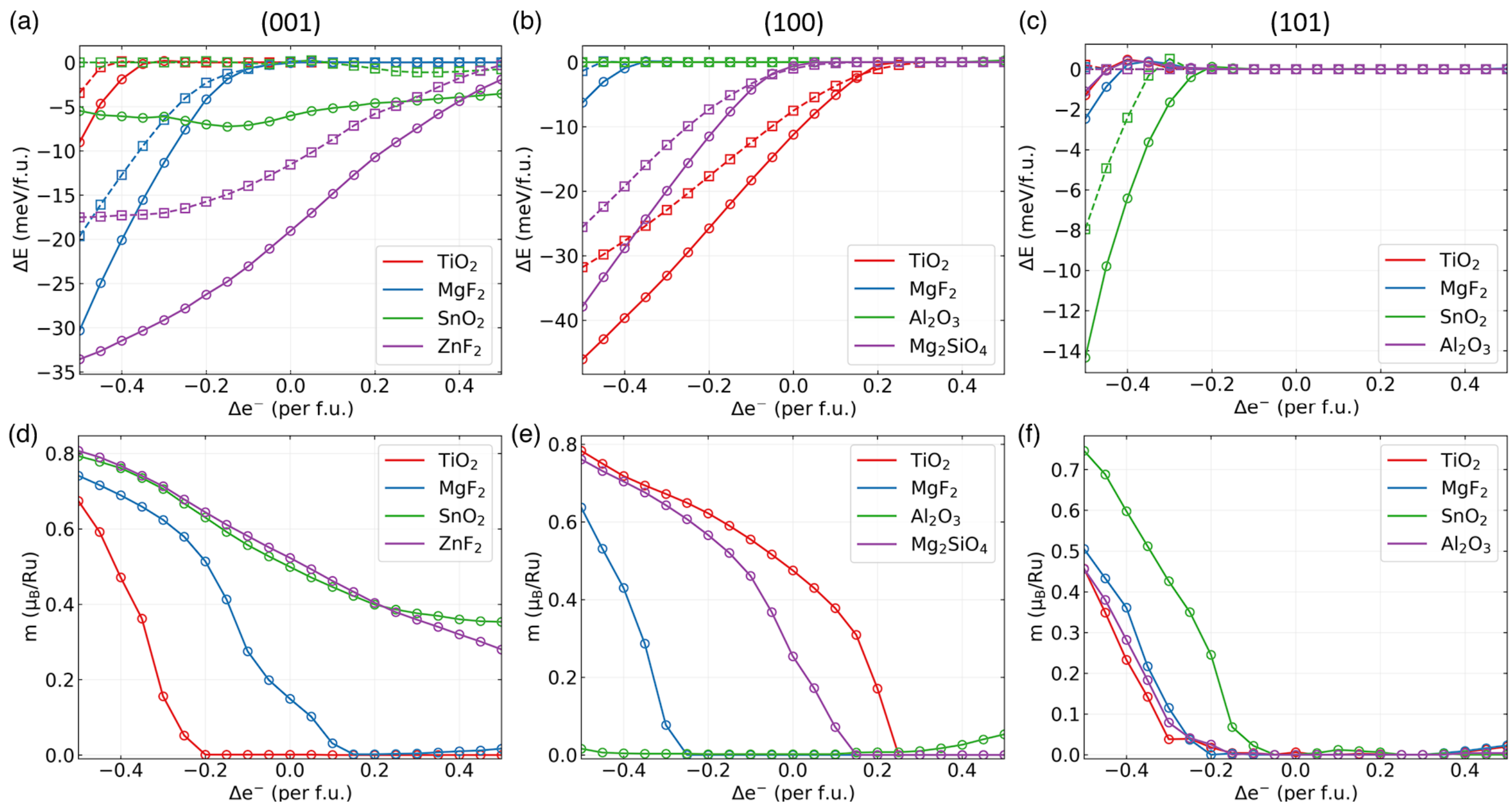

**Figure 6:** Doping dependence of the magnetic ground state for epitaxially strained $RuO_2$. All calculations were performed without a Hubbard U. (a-c) Energy difference relative to the nonmagnetic phase for the altermagnetic (solid lines) and ferromagnetic (dashed lines) phases as functions of doping for the (001), (100), and (101) orientations, respectively. (d-f) The corresponding Ru magnetic moments of the altermagnetic phase. Negative (positive) sign of $\Delta e^-$ denotes the hole (electron) doping.

To further investigate the effect of the electronic interactions, we calculate the spin susceptibility within the random-phase approximation (RPA) [117, 118 ], using the following Hubbard-Hund interaction:

$$H_{\mathrm{int}} = \sum_{i\alpha}\Bigg[U\sum_{l} n_{i\alpha l\uparrow}n_{i\alpha l\downarrow} + U'\sum_{l<l'} n_{i\alpha l}n_{i\alpha l'} + J\sum_{\substack{l<l'\\ \sigma\sigma'}} c^\dagger_{i\alpha l\sigma}c^\dagger_{i\alpha l'\sigma'}c_{i\alpha l\sigma'}c_{i\alpha l'\sigma} + J'\sum_{l\neq l'} c^\dagger_{i\alpha l\uparrow}c^\dagger_{i\alpha l\downarrow}c_{i\alpha l'\downarrow}c_{i\alpha l'\uparrow}\Bigg],$$

where $c_{i\alpha l\sigma}$ is the electron annihilation operator at site $(i\alpha)$ with orbital $l$ and spin $\sigma$, $n_{i\alpha l\sigma}$ is the density operator. The $U, U', J,$ and $J'$ denote the intra-orbital and inter-orbital Hubbard interaction, the Hund's coupling and pair-hopping terms, respectively. The corresponding RPA spin susceptibility matrix is then given by

$$\chi^s_{RPA}(\boldsymbol{q}) = \chi_0(\boldsymbol{q})[\mathbb{1} - U^s\chi_0(\boldsymbol{q})]^{-1}\,,$$

where $U^s$ represents the interaction matrix for the spin channel, with element

$$[U^s]^{l_1l_2}_{l_3l_4} = \begin{cases} U, & l_1 = l_2 = l_3 = l_4 \\ U', & l_1 = l_3 \neq l_2 = l_4 \\ J, & l_1 = l_2 \neq l_3 = l_4 \\ J', & l_1 = l_4 \neq l_2 = l_3 \end{cases}$$

We impose the rotationally invariant Kanamori relations, $U' = U - 2J$, and $J' = J$, leaving only two independent interaction parameters, $U$ and $J$. The ferromagnetic and altermagnetic RPA susceptibility is then given by

$$\chi^{\tau_M}_{\mathrm{RPA}}(\boldsymbol{q}) = \sum_{\mu_1\mu_2\mu_3\mu_4} (\tau_M)_{\mu_1\mu_2}\,(\tau_M)^*_{\mu_3\mu_4}\,[\chi^s_{\mathrm{RPA}}(\boldsymbol{q})]^{\mu_1\mu_2}_{\mu_3\mu_4}\,.$$

Figure 5 presents the RPA susceptibilities calculated with $U = 0.5\,eV$ and $J/U = 0.15$ for bulk $RuO_2$ and the epitaxial strain conditions corresponding to different substrates and orientations. For bulk $RuO_2$, $RuO_2$(001)/$TiO_2$(001), $RuO_2$(100)/$MgF_2$(100), $RuO_2$(100)/$TiO_2$(100), and $RuO_2$(001)/$ZnF_2$(001), the susceptibility peaks occur at the Γ point, with $\chi^{\mathrm{AM}}_{\mathrm{RPA}}(\Gamma) > \chi^{\mathrm{FM}}_{\mathrm{RPA}}(\Gamma)$. In contrast, for $RuO_2$(100)/$SrTiO_3$(100), including interactions shifts the altermagnetic susceptibility maxima from Γ in bare response to finite wave vectors along the $\Gamma - Z$ path, at $q_z = 0.01\frac{2\pi}{c}$, suggesting interaction-induced spin-density-wave instability. This is qualitatively consistent with the experimentally reported incommensurate spin-density-wave order in $RuO_2$(100)/$SrTiO_3$(100), although determining the ordering wave vector requires further investigation. For $RuO_2$(001)/$SnO_2$(001), whose bare susceptibilities already exhibit finite-wave-vector maxima, the RPA peaks shift further along the $\Gamma - Z$ path to $q_z = 0.113\frac{2\pi}{c}$. The peak for the $RuO_2$(100)/$Mg_2SiO_4$(010) stays at the same wave vector $q_z = 0.1017\frac{2\pi}{c}$ in $\Gamma - Z$ path. The results with different $U$ and $J$ parameters are discussed in Sec. V of Supplemental Material [104].

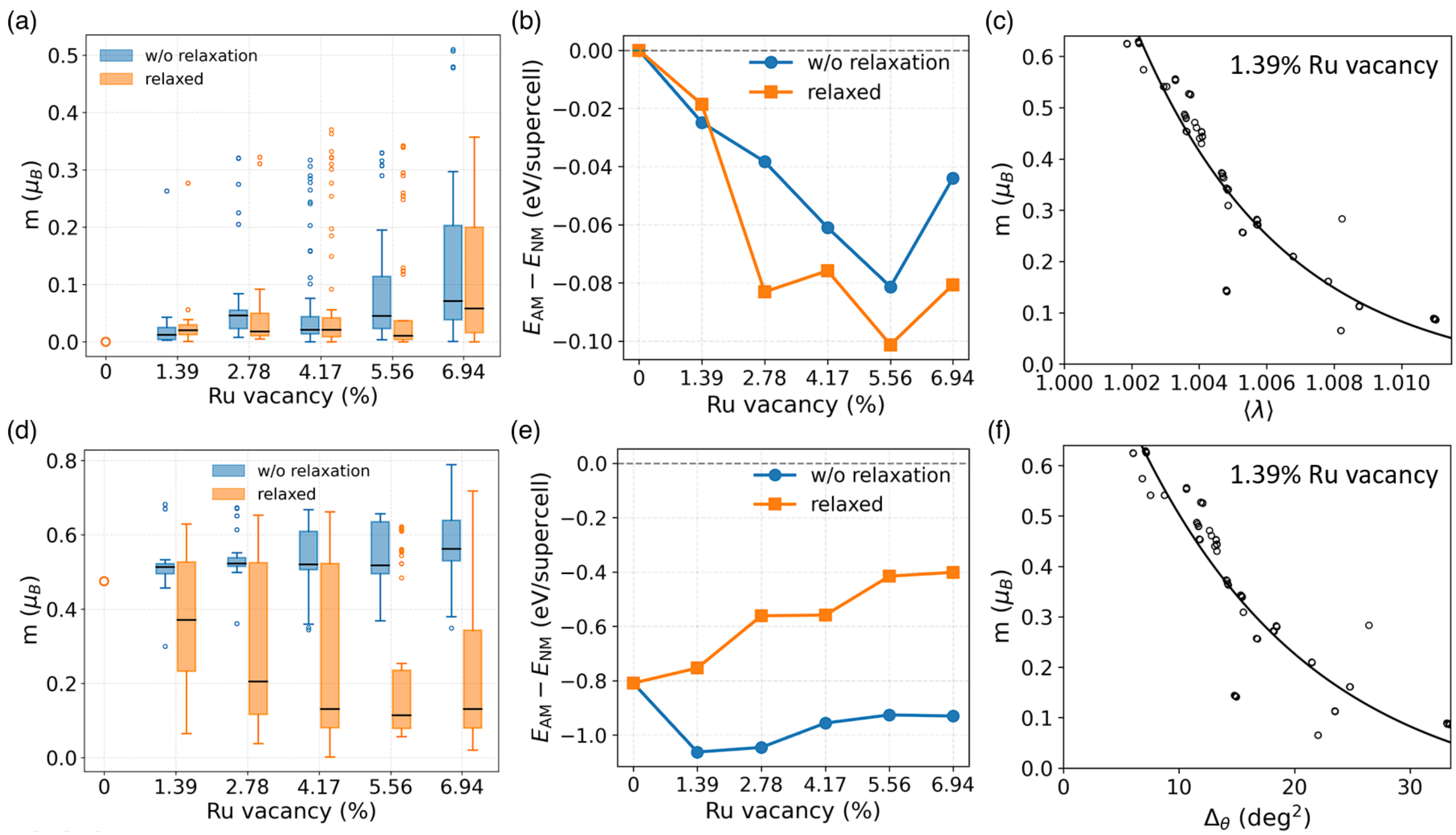

**Figure 7:** The Ru vacancy dependence of the magnetism for bulk $RuO_2$ and $RuO_2$(100)/$TiO_2$(100). (a,b) Distributions of the Ru magnetic moments and the energy difference $E_{AM} - E_{NM}$, respectively, as functions of the Ru-vacancy concentration in bulk $RuO_2$ supercells. (d,e) Corresponding results for $RuO_2$(100)/$TiO_2$(100) supercells. Blue and orange color denote structures without and with atomic relaxation after introducing the vacancies, respectively. In each box plot, the central line represents the median, the box spans the interquartile range, the whiskers show the overall range, and individual points indicate outliers. (c,f) Ru magnetic moments as functions of the mean quadratic elongation (c) and cis-angle variance (f) of the $RuO_6$ octahedra at a 1.39% Ru-vacancy in $RuO_2$(100)/$TiO_2$(100).

### D. Effects of Vacancies and Doping

As discussed above, strain shifts the density-of-states peak in $RuO_2$ towards the Fermi level, stabilizing altermagnetism. Hole doping provides an alternative route by lowering the Fermi level toward this peak [50,69,105,119]. Previous DFT+U studies showed that hole doping can stabilize an altermagnetic ground state in bulk $RuO_2$ and proposed that a 10% Ru-vacancy concentration could produce a Ru moment of approximately 0.2 $\mu_B$ [69]. In our calculations without Hubbard $U$, 0.5 holes per Ru are needed to stabilize the altermagnetic state, yielding a moment of 0.37 $\mu_B$ (Supplemental Fig. S1 [104]). Including $U$ reduces the required hole concentration to 0.2 holes per Ru and increases the moment to 0.6 $\mu_B$. Thus, our calculations indicate that stabilizing bulk altermagnetism requires substantial hole doping, although $U$ lowers the required concentration.

We next examine hole doping in $RuO_2$ films oriented along the (001), (100), and (101) directions under substrate-imposed epitaxial strain, without including Hubbard $U$. As shown in Fig. 6, hole doping increases both the energy separation between the altermagnetic and nonmagnetic states and the Ru moment for all structures considered. In $RuO_2$(100)/$TiO_2$(100), and $RuO_2$(001)/$ZnF_2$(001), doping with 0.2 holes per Ru raises the moment to approximately 0.6 $\mu_B$ and increases the energy separation between the two states. Doping also substantially enhances the moments in $RuO_2$(001)/$MgF_2$(001), and $RuO_2$(100)/$Mg_2SiO_4$(010), where the undoped moments are only approximately 0.15 and 0.25 $\mu_B$, respectively. At 0.2 holes per Ru, the moments in both films exceed 0.5 $\mu_B$. By comparison, the (101)-oriented films require more than 0.2 holes per Ru to reach comparable moments, regardless of substrate.

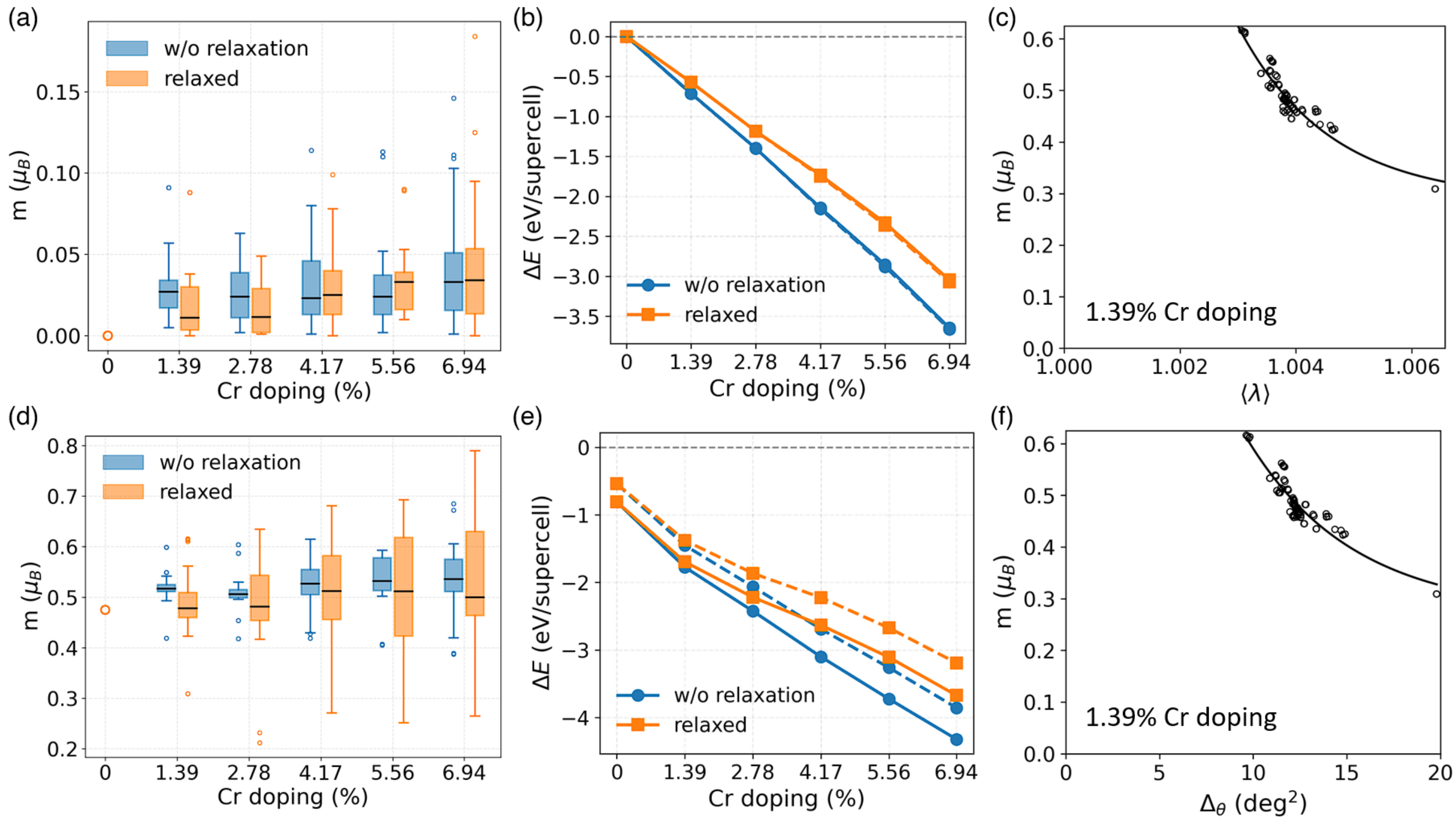

**Figure 8:** The Cr doping dependence of the magnetism for bulk $RuO_2$ and $RuO_2$(100)/$TiO_2$(100). (a,b) Distributions of the Ru magnetic moments and the energy difference $E_{AM} - E_{NM}$ (solid line) and $E_{FM} - E_{NM}$ (dashed line), as functions of the Cr doping concentration in bulk $RuO_2$ supercells. (d,e) Corresponding results for $RuO_2$(100)/$TiO_2$(100) supercells. Blue and orange color denote structures without and with atomic relaxation after introducing the Cr doping, respectively. In each box plot, the central line represents the median, the box spans the interquartile range, the whiskers show the overall range, and individual points indicate outliers. (c,f) Ru magnetic moments as functions of the mean quadratic elongation (c) and cis-angle variance (f) of the $RuO_6$ octahedra at a 1.39% Cr-doping in $RuO_2$(100)/$TiO_2$(100).

In the previous analysis, hole doping was modeled by uniformly distributing the compensating charge throughout the cell, which differs from realistic experimental conditions. To obtain a more experimentally relevant description, we therefore investigate hole doping introduced through Ru vacancies and substitutional Cr and Mn dopants. We consider $3 \times 3 \times 4$ supercells containing 72 formula units of $RuO_2$ for both unstrained bulk $RuO_2$ and strained $RuO_2$(100)/$TiO_2$(100). Defective structures are generated by either removing Ru atoms or replacing them with dopant atoms. The positions of the defects are chosen to maximize their separation and thereby minimize defect-defect interactions.

Figure 7(a) shows the distribution of local Ru magnetic moments as a function of Ru-vacancy concentration in bulk $RuO_2$. Although the moments generally increase with increasing vacancy concentration, most Ru atoms have relatively small moments, with median values below 0.1 $\mu_B$. The altermagnetic state is energetically favorable relative to the nonmagnetic state, but its energy gain remains very small [Fig. 7(b)].

A different behavior is observed for strained $RuO_2$(100)/$TiO_2$(100). Before atomic relaxation, introducing Ru vacancies increases the Ru magnetic moments and enhances the energetic preference for the altermagnetic state [Figs. 7(d,e)]. After relaxation, however, the magnetic-moment distribution shifts toward smaller values as the vacancy concentration increases, with the median falling below 0.2 $\mu_B$ at a vacancy concentration of 6.94 %. Consistently, the energetic stabilization of the altermagnetic state decreases with increasing Ru-vacancy concentration. This suppression can be attributed to vacancy-induced structural distortions. In pristine $RuO_2$(100)/$TiO_2$(100), the mean quadratic elongation and cis-angle variance of the $RuO_6$ octahedra are approximately 1.0039 and 12.679 deg$^2$, respectively. At a Ru-vacancy concentration of 1.39%, both distortion measures increase substantially for many $RuO_6$ octahedra [Figs. 7(c,f)]. The observed trend indicates that more strongly distorted octahedra generally carry smaller Ru magnetic moments, agreeing with discussion in sec. III B. Thus, Ru vacancies are not an effective means of stabilizing altermagnetism in $RuO_2$ after structural relaxation, they can instead weaken the strain-stabilized altermagnetic state.

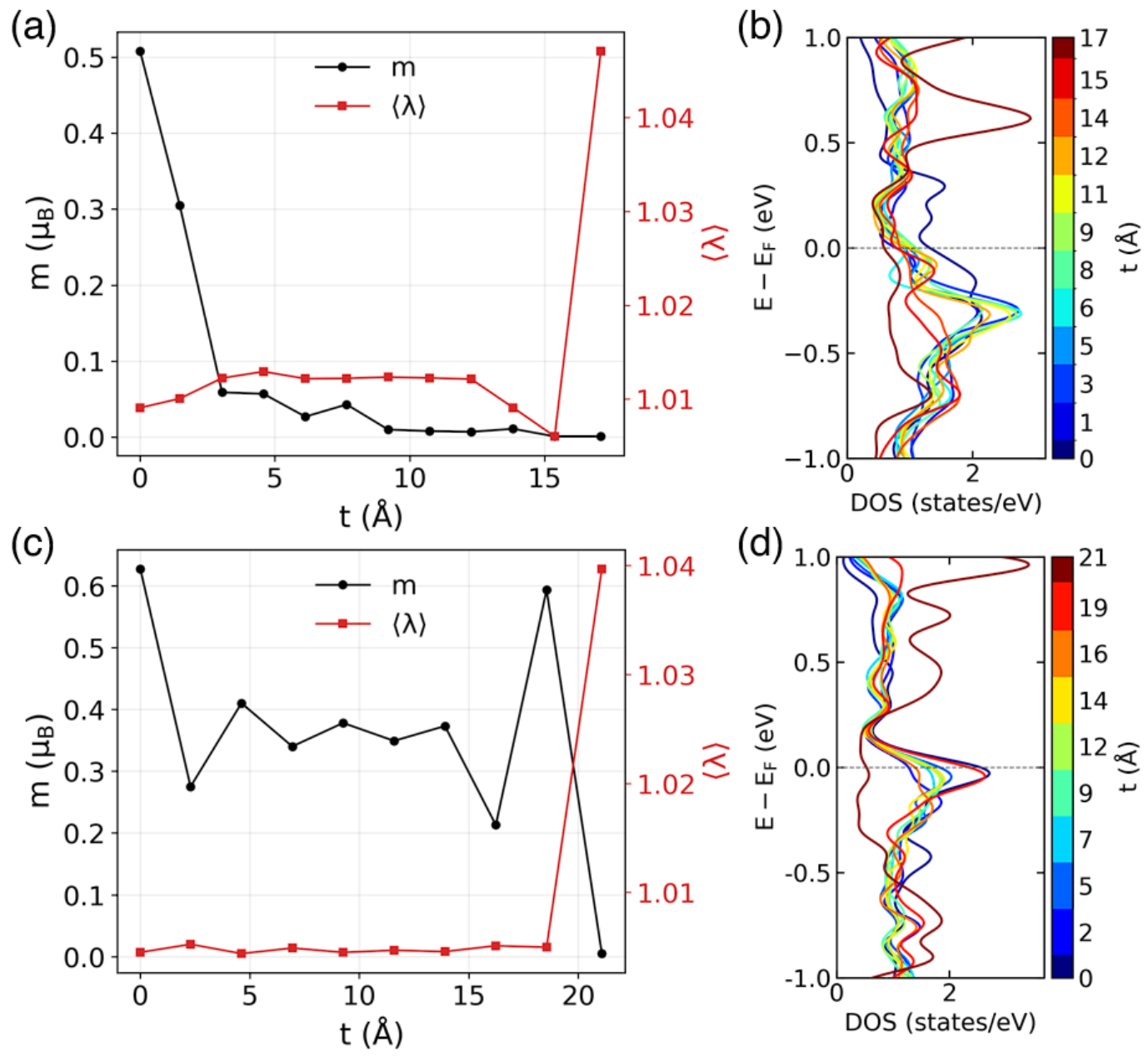

**Figure 9:** (a,c) Local Ru magnetic moments and quadratic elongations of the $RuO_6$ octahedra as functions of distance from the $RuO_2/TiO_2$ interface for the (001)- and (100)-oriented heterostructures, respectively. (b,d) Corresponding layer-projected densities of states (DOS) calculated in the nonmagnetic state.

We next examine the effect of Cr doping. Figure 8(a) shows the distribution of local Ru magnetic moments as a function of Cr concentration in bulk $RuO_2$. Compared with the Ru-vacancy case, Cr substitution produces only a small enhancement of the Ru moments, with median values remaining below 0.1 $\mu_B$. In contrast, the Cr dopants develop large local moments of approximately 2.065 $\mu_B$ . Although the energy difference between the nonmagnetic and altermagnetic states is significant [Fig. 8(b)], the magnetism is dominated by the Cr moments rather than by an enhancement of the Ru moments. Furthermore, the ferromagnetic state is energetically favored over the altermagnetic state in bulk $RuO_2$, with an energy difference of 23.6 meV per supercell at a Cr concentration of 6.94 %.

For strained $RuO_2(100)/TiO_2(100)$, Cr substitution increases the Ru magnetic moments and enhances the energetic preference for the altermagnetic state before atomic relaxation [Figs. 8(d,e)]. After relaxation, the distribution of Ru moments broadens, while its median remains approximately 0.5 $\mu_B$. For both relaxed and unrelaxed structures, the energetic stabilization of the altermagnetic state relative to the nonmagnetic state increases with Cr concentration. As shown in Figs. 8(c,f), Cr substitution produces substantially smaller local distortions than Ru vacancies. Consequently, the reduction of the Ru moments following atomic relaxation is much less pronounced than in the Ru-vacancy case. Moreover, unlike unstrained bulk $RuO_2$, the altermagnetic state remains lower in energy than the ferromagnetic state. These results indicate that Cr doping can further stabilize the strain-induced altermagnetic phase in $RuO_2(100)/TiO_2(100)$. The Mn doping shows similar results as discussed in Sec. VII of Supplemental Material [104].

Overall, these results demonstrate that neither Ru vacancies nor chemical doping alone are sufficient to stabilize altermagnetism in bulk $RuO_2$. Strain remains the primary factor driving the stabilization of the altermagnetic phase. Chemical doping can further enhance strain-induced altermagnetism, whereas Ru vacancies tend to weaken it.

### E. Effects of Interface

Another possible mechanism for enhancing magnetism is electrostatic doping induced by charge transfer across the $RuO_2$ and substrate interface. To investigate this effect, we construct symmetric $RuO_2/TiO_2/RuO_2$ heterostructures containing a four-layer $TiO_2$ slab enclosed by five-layer $RuO_2$ slabs on both sides (see Sec. VIII of Supplemental Material for the side view [104]). We consider both the (001) and (100) orientations, which were found to be nonmagnetic and altermagnetic, respectively, in the corresponding strained-bulk calculations. The central one layer of $TiO_2$ is constrained to its experimental bulk geometry, while all remaining atoms are allowed to relax. The surface terminations are chosen to preserve complete $RuO_6$ octahedra.

As shown in Figs. 9(a,c), the Ru moments are enhanced in the first $RuO_2$ layer adjacent to the $TiO_2$ interface. This layer also has the largest peak in its projected density of states at the Fermi level [Figs. 9(b,d)], consistent with hole doping caused by interfacial charge transfer. Moving away from the interface, the Ru moments gradually approach the values calculated for the corresponding epitaxially strained bulk phases.

The behavior near the free surface is different. The surface $RuO_6$ octahedra are strongly distorted, as indicated by their large quadratic elongation. This distortion suppresses the density of states at the Fermi level, and the surface Ru moments vanish. In the layer just below the surface, however, the density of states at the Fermi level increases substantially, and the Ru moments are enhanced. Although epitaxially strained bulk $RuO_2(001)/TiO_2(001)$ is nonmagnetic in our calculations, local magnetism near the interface could contribute to the experimental magnetic signatures reported for this heterostructure.

## IV. DISCUSSION

Our results show that the magnetic behavior of $RuO_2$ depends on its structural and chemical environment. In our calculations, unstrained stoichiometric bulk $RuO_2$ favors a nonmagnetic state, whereas selected epitaxial strains stabilize ordered Ru moments. Doping and interfaces can further modify the strength and spatial distribution of magnetism. This picture provides a framework for interpreting why magnetic signatures differ between bulk crystals and films. A magnetic or electrical signal from a film alone does not establish altermagnetic order throughout the sample. That interpretation requires identifying the Ru magnetic state and determining its spatial extent.

The case against a sizeable, homogeneous magnetic moment in bulk $RuO_2$ has become substantial. The neutron and resonant X-ray studies that motivated the original antiferromagnetic picture reported weak reflections interpreted as magnetic [28,29]. Subsequent muon measurements found no spontaneous internal field in crystals [30], and the combined muon and neutron study set a much smaller moment bound and attributed the weak neutron intensity to multiple scattering [31]. Single-crystal diffraction, lattice-dynamics and Mössbauer measurements, and bulk transport likewise place constraints on static order [32,48,49,53,54]. Particularly informative are quantum oscillations, which map a multisheet Fermi surface consistent with a nonmagnetic metal rather than the predicted ordered bulk state [33]. The absence of the expected bulk spin splitting in photoemission adds a different electronic-structure test [51,52,85]. Each method has its own sensitivity and sampling depth, but their convergence is consistent with our nonmagnetic bulk ground state. It does not exclude small, disordered moments, a surface contribution, or magnetism created by a different strain or defect environment.

The theoretical literature also points to proximity to magnetic instability rather than unconditionally ordered bulk material. The calculations found that the interaction strength consistent with stoichiometric $RuO_2$ is below the threshold for magnetic order but suggested that Ru vacancies could push the system across it [50]. Other calculations emphasize the sensitivity of competing instabilities to interactions, doping, and the Fermi surface [69,105,114,115,120−123]. Our result that an altermagnetic state emerges only beyond a finite Hubbard correction agrees with the first conclusion, although the threshold is method dependent. DFT can favor moment formation in itinerant metals, while the appropriate interaction parameters remain experimentally unconstrained. Therefore, a magnetic DFT solution alone does not establish the ground state. The nonmagnetic PBE result likewise cannot rule out correlation-driven or fluctuating phases. The susceptibility calculations identify states whose magnetic response is enhanced by strain. An enhanced susceptibility, including a peak at finite wave vector, signals a magnetic tendency but does not determine a transition temperature or establish long-range order.

Epitaxial strain offers a coherent explanation for some differences between bulk crystals and films, but orientation and actual strain retention matter. Our calculations stabilize the compensated ordered state for $RuO_2$(100)/$TiO_2$(100) and predict favorable conditions for selected (001) films on $MgF_2$, $SnO_2$, and $ZnF_2$, whereas $RuO_2$(001)/$TiO_2$(001) remains nonmagnetic without an added Hubbard term. This orientation dependence cautions against interpreting all $TiO_2$-supported films as equivalent. In strained ultrathin $RuO_2$ films, one study reported an anomalous Hall response that weakened as the strain relaxed [37], while another mapped a thickness-dependent phase diagram and identified a polar metallic magnetic phase in films about 2 nm thick [34]. These observations support a strong role for the epitaxial environment, although Hall transport alone cannot determine the microscopic Ru ordering pattern. The optical and vibrational comparison of films on several $TiO_2$ orientations also documents sensitivity to strain and electronic structure without by itself settling the magnetic ground state [105]. Our uniform-strain model is most directly applicable where coherency is verified; strain gradients, dislocations, reconstruction, and partial relaxation in real films require separate treatment [109,112,124].

Several thin film results present direct tests and unresolved discrepancies for the present model. For $RuO_2$(100)/$Al_2O_3$(0001), experiments report altermagnetic signatures, yet our imposed coherent geometry favors the nonmagnetic state at $U = 0$. The very large nominal mismatch makes a uniformly strained film unlikely, so the experimental atomic structure may differ substantially from the assumed cell [36,112]. For (101)-oriented $RuO_2$, our strained-bulk calculations also favor a nonmagnetic state on the substrates considered, whereas single-variant $RuO_2$ (101) films on sapphire have been interpreted as magnetically ordered using structural analysis, X-ray magnetic linear dichroism, and bilayer magnetoresistance [58]. That study highlights the possible importance of variant selection, interface registry, and annealing, none of which is fully represented by the present bulk-like strain model. We therefore regard these cases as genuine disagreements requiring film-specific structural models and independent depth-resolved tests, rather than as evidence that either the calculation or the experiment is generically invalid.

The (110) orientation raises an additional distinction between induced moments and a compensated altermagnetic state. In our coherently strained model, the two Ru sites acquire unequal moments, producing a ferrimagnetic solution. This prediction cannot be equated directly with reports of an altermagnetic polar phase in extremely thin $RuO_2$/$TiO_2$ (110) films [34], because the actual film includes surfaces, a substrate interface, and structural changes absent from the strained-bulk model. Similarly, our predicted net moment has not been established by the measurements discussed here. A quantitative comparison would require the measured lattice and oxygen positions, layer-resolved calculations, and a search for the net moment with sufficient sensitivity. Depth-sensitive muon measurements have also detected inhomogeneous static magnetism concentrated near film surfaces [86], underscoring the need to distinguish surface or interfacial order from magnetism throughout the film.

Transport reports require similar care. Spin-splitting torques, spin-to-charge conversion, magnetoresistance, and field-free switching have been attributed to altermagnetic $RuO_2$ in various heterostructures [20,36,61,63–66,87,88,93,101]. Other experiments find no corresponding transport spin-splitting signature or instead identify contributions associated with interfaces and conventional conversion mechanisms [35,67,68]. A torque measures the response of the complete multilayer, including interfaces, adjacent magnetic layers, current paths, and possible spin Hall contributions; it is not a direct measurement of a spatially uniform Ru moment. The contrasting reports can therefore coexist if growth conditions produce different local magnetic regions or if distinct mechanisms dominate in different stacks. Our results suggest that substrate orientation, coherent strain, and the depth profile of Ru moments should be measured alongside torque anisotropy before assigning a response to altermagnetic band splitting.

The anomalous Hall effect and photoemission lead to a related caution. Hall signals in strained films [37,97] and spin-sensitive photoemission interpreted as broken time-reversal symmetry [55,56] warrant serious consideration, yet other angle-resolved measurements on films and crystals do not observe the predicted bulk spin splitting [51,95]. Surface electronic bands, film thickness, strain, and spatially nonuniform order could make these probes sample physically different states. Hall response can also arise from mechanisms other than a homogeneous Ru altermagnet, particularly in doped or reconstructed samples [119, 125 ]. $Ti^{+3}$ states arising from interfacial reduction or interdiffusion can contribute an additional Hall signal that must be separated from the strain-driven contribution [126].

Our susceptibility analysis provides one possible route to understand reports of nonuniform order without assuming that every strained film is a collinear altermagnet. For several substrate-imposed structures, the strongest response occurs at the collinear ordering wave vector, whereas for others it moves to a finite wave vector. In addition, interactions can also shift the leading instability. In particular, the finite-wave-vector tendency calculated for $RuO_2$(100)/$SrTiO_3$(100) is qualitatively compatible with the incommensurate spin structure inferred from spin-wave-coupled torque measurements [62]. Those measurements estimate a real-space wavelength, whereas our susceptibility peak is derived from an idealized electronic model. Their agreement is therefore suggestive, not a quantitative identification of the observed modulation. Direct scattering or spatially resolved magnetic imaging on structurally characterized films would test whether the proposed spin-density wave, compensated collinear order, or another texture forms.

Finally, nominal hole doping is not equivalent to introducing a chemical defect. Earlier calculations proposed that carrier removal or Ru vacancies could stabilize moments [50,69,105], and our uniform-charge calculations indeed find that holes strengthen the magnetic tendency. Relaxed Ru vacancies, however, distort the neighboring octahedra and substantially weaken the ordered state in strained $RuO_2$(100)/$TiO_2$(100). This difference is a result of local structure, not a contradiction in electron counting. Cr substitution also produces large dopant moments but little Ru magnetization in unstrained bulk. A magnetic response from Cr-doped material is therefore insufficient to identify altermagnetic Ru order. Reports of competing magnetic interactions or impurity-related phases in Cr-doped samples reinforce this distinction [70,119]. Under favorable epitaxial strain, by contrast, Cr and Mn substitution can enhance Ru moments while preserving the energetic preference for the compensated state in our models. The experimentally decisive quantities are the Ru-specific moments, dopant distribution, local coordination, and spatial extent of order, rather than the nominal dopant concentration alone.

Interfaces offer a final route by which an apparently contradictory set of measurements may reflect different parts of the same sample. In our $RuO_2$/$TiO_2$ slabs, interfacial charge transfer enhances moments in the adjacent $RuO_2$ layer even where the corresponding strained bulk model is nonmagnetic, while strong surface relaxation suppresses moments in the outermost layer. The resulting nonmonotonic profile is broadly compatible with proposals of surface or interface magnetism [71–73], but our ideal slabs do not reproduce every possible termination, oxygen defect, or film thickness. Future comparisons should combine crystallographically resolved strain and oxygen occupancy, element-specific and depth-sensitive magnetic probes, and electronic measurements on the same specimens. Such a program would establish whether a reported signal belongs to bulk-like Ru order, a reconstructed film layer, a dopant-rich region, or a competing finite-wave-vector phase.

## V. CONCLUSIONS

We have investigated how structural and chemical changes affect magnetism in $RuO_2$ using first-principles calculations and spin-susceptibility analysis. Without an empirical Hubbard correction, unstrained bulk $RuO_2$ is nonmagnetic in our calculations. Epitaxial strain can stabilize magnetic order on the Ru sublattices, with a strong dependence on substrate and film orientation. Altermagnetic ground states are predicted for several (001)-oriented and selected (100)-oriented configurations, whereas the (101) configurations examined remain nonmagnetic.

Spin-susceptibility analysis reveals the electronic origin of this strain dependence. Strain shifts a density-of-states peak, located approximately 0.6 eV below the Fermi level in bulk $RuO_2$, toward the Fermi level, enhancing the susceptibility and the tendency toward magnetic ordering. The dominant response does not always correspond to the collinear altermagnetic state: under some substrate-imposed strain conditions, a susceptibility peak at a finite wave vector suggests a competing spin-density-wave instability. This competition provides a possible basis for the diverse magnetic responses reported in epitaxial $RuO_2$ films.

The effects of chemical modification depend strongly on the structural environment. Uniform hole doping enhances Ru magnetic moments, but Ru vacancies and substitutional dopants produce additional local structural and chemical changes. Neither Ru vacancies nor Cr or Mn substitution stabilizes robust collinear altermagnetic order in unstrained bulk $RuO_2$ within our calculations. Under epitaxial strain, however, Cr and Mn substitution can reinforce Ru magnetism, whereas relaxation around Ru vacancies can weaken it through local octahedral distortions. $RuO_2/TiO_2$ interfaces further modify the magnetic state, producing layer-dependent moments through charge transfer and structural relaxation.

Overall, these results identify epitaxial strain as the principal route to stabilizing Ru magnetism among the conditions examined and show how chemical modification and interfaces alter its strength and spatial distribution. They also identify substrate-orientation combinations for experimental tests of strain-stabilized altermagnetism. Establishing whether this order is realized in experimental films requires measurements that resolve the spatial distribution of Ru moments and distinguish long-range altermagnetic order from competing magnetic states and localized interfacial magnetism.

## ACKNOWLEDGMENTS

The authors thank Prof. Cheng Song, Prof. Bharat Jalan, and Hazem Abdelhafiz for useful discussions. This work was supported by the National Science Foundation (NSF), Division of Materials Research, under Award No. DMR-2316665, and by the NSF Materials Research Science and Engineering Center (MRSEC) at the University of Nebraska-Lincoln under Award No. 2614708. Computational resources were provided by the Holland Computing Center at the University of Nebraska-Lincoln.

## DATA AVAILABILITY

The data that support the findings of this paper are available from the authors upon reasonable request.

1. V. Baltz, A. Manchon, M. Tsoi, T. Moriyama, T. Ono, and Y. Tserkovnyak, Antiferromagnetic spintronics, Rev. Mod. Phys. **90**, 015005 (2018).
2. T. Jungwirth, J. Sinova, A. Manchon, X. Marti, J. Wunderlich, and C. Felser, The multiple directions of antiferromagnetic spintronics, Nat. Phys. **14**, 200 (2018).
3. S. Hayami, Y. Yanagi, and H. Kusunose, Momentum-dependent spin splitting by collinear antiferromagnetic ordering, J. Phys. Soc. Jpn. **88**, 123702 (2019).
4. L.-D. Yuan, Z. Wang, J.-W. Luo, E. I. Rashba, and A. Zunger, Giant momentum-dependent spin splitting in centrosymmetric low-Z antiferromagnets, Phys. Rev. B **102**, 014422 (2020).
5. L.-D. Yuan, Z. Wang, J.-W. Luo, and A. Zunger, Prediction of low-Z collinear and noncollinear antiferromagnetic compounds having momentum-dependent spin splitting even without spin-orbit coupling, Phys. Rev. Materials **5**, 014409 (2021).
6. L. Šmejkal, J. Sinova, and T. Jungwirth, Beyond conventional ferromagnetism and antiferromagnetism: A phase with nonrelativistic spin and crystal rotation symmetry, Phys. Rev. X **12**, 031042 (2022).
7. L. Šmejkal, J. Sinova, and T. Jungwirth, Emerging research landscape of altermagnetism, Phys. Rev. X **12**, 040501 (2022).
8. M. Naka, S. Hayami, H. Kusunose, Y. Yanagi, Y. Motome, and H. Seo, Spin current generation in organic antiferromagnets, Nat. Commun. **10**, 4305 (2019).
9. R. González-Hernández, L. Šmejkal, K. Výborný, Y. Yahagi, J. Sinova, T. Jungwirth, J. Železný, Efficient electrical spin splitter based on nonrelativistic collinear antiferromagnetism, Phys. Rev. Lett. **126**, 127701 (2021).
10. D.-F. Shao, S.-H. Zhang, M. Li, C.-B. Eom, and E. Y. Tsymbal, Spin-neutral currents for spintronics, Nat. Commun. **12**, 7061 (2021).
11. R. Takagi, R. Hirakida, Y. Settai, R. Oiwa, H. Takagi, A. Kitaori, K. Yamauchi, H. Inoue, J. Yamaura, D. Nishio-Hamane, S. Itoh, S. Aji, H. Saito, T. Nakajima, T. Nomoto, R. Arita, and S. Seki, Spontaneous Hall effect induced by collinear antiferromagnetic order at room temperature, Nat. Mater. **24**, 63 (2025).
12. X.-J. Yi, Y. Mao, X. Lu, and Q.-F. Sun, Spin splitting Nernst effect in altermagnets, Phys. Rev. B **111**, 035423 (2025).
13. K. Sourounis and A. Manchon, Efficient generation of spin currents in altermagnets via magnon drag, Phys. Rev. B **111**, 134448 (2025).
14. K. Samanta, D.-F. Shao, and E. Y. Tsymbal, Spin filtering with insulating altermagnets, Nano Lett. **25**, 3150 (2025).
15. Y.-F. Sun, Y. Mao, Y.-C. Zhuang, and Q.-F. Sun, Tunneling magnetoresistance effect in altermagnets, Phys. Rev. B **112**, 094411 (2025).
16. X.-P. Zhang, W. Feng, R.-W. Zhang, X. Fan, X. Wang, and Y. Yao, Theory of anisotropic magnetoresistance in altermagnets and its applications, Phys. Rev. Lett. **135**, 266706 (2025).
17. M. Ezawa, Tunneling magnetoresistance in a junction made of $X$-wave magnets with $X = p, d, f, g, i$, Phys. Rev. B **113**, 155303 (2026).
18. R. D. González Betancourt, J. Zubáč, K. Geishendorf, P. Ritzinger, B. Růžičková, T. Kotte, J. Železný, K. Olejník, G. Springholz, B. Büchner, A. Thomas, K. Výborný, T. Jungwirth, H. Reichlová, and D. Kriegner, Anisotropic magnetoresistance in altermagnetic MnTe, npj Spintronics **2**, 45 (2024).

19. S. Noh, G.-H. Kim, J. Lee, H. Jung, U. Seo, G. So, J. Lee, S. Lee, M. Park, S. Yang, Y. S. Oh, H. Jin, C. Sohn, and J.-W. Yoo, Tunneling magnetoresistance in altermagnetic $RuO_2$-based magnetic tunnel junctions, Phys. Rev. Lett. **134**, 246703 (2025).
20. S. Karube, T. Tanaka, D. Sugawara, N. Kadoguchi, M. Kohda, and J. Nitta, Observation of spin-splitter torque in collinear antiferromagnetic $RuO_2$, Phys. Rev. Lett. **129**, 137201 (2022).
21. H. Mavani, K. Huang, K. Samanta, and E. Y. Tsymbal, Two-dimensional antiferromagnets with nonrelativistic spin splitting switchable by electric polarization, Phys. Rev. B **112**, L060401 (2025).
22. C. Song, H. Bai, Z. Zhou, L. Han, H. Reichlova, J. H. Dil, J. Liu, X. Chen, and F. Pan, Altermagnets as a new class of functional materials, Nat. Rev. Mater. **10**, 473 (2025).
23. T. Jungwirth, J. Sinova, R. M. Fernandes, Q. Liu, H. Watanabe, S. Murakami, S. Nakatsuji, and L. Šmejkal, Symmetry, microscopy and spectroscopy signatures of altermagnetism, Nature **649**, 837 (2026).
24. L. Šmejkal, R. González-Hernández, T. Jungwirth, and J. Sinova, Crystal time-reversal symmetry breaking and spontaneous Hall effect in collinear antiferromagnets, Sci. Adv. **6**, eaaz8809 (2020).
25. W. D. Ryden and A. W. Lawson, Magnetic susceptibility of $IrO_2$ and $RuO_2$, J. Chem. Phys. **52**, 6058 (1970).
26. H. Mukuda, K. Ishida, Y. Kitaoka, K. Asayama, R. Kanno, and M. Takano, Spin fluctuations in the ruthenium oxides $RuO_2$, $SrRuO_3$, $CaRuO_3$, and $Sr_2RuO_4$ probed by Ru NMR, Phys. Rev. B **60**, 12279 (1999).
27. Y.-X. Li, Y. Chen, L. Pan, S. Li, S.-B. Zhang, and H.-Z. Lu, Exploration of altermagnetism in $RuO_2$, Sci. China Phys. Mech. Astron. **69**, 257001 (2026).
28. T. Berlijn, P. C. Snijders, O. Delaire, H.-D. Zhou, T. A. Maier, H.-B. Cao, S.-X. Chi, M. Matsuda, Y. Wang, M. R. Koehler, P. R. C. Kent, and H. H. Weitering, Itinerant antiferromagnetism in $RuO_2$, Phys. Rev. Lett. **118**, 077201 (2017).
29. Z. H. Zhu, J. Strempfer, R. R. Rao, C. A. Occhialini, J. Pelliciari, Y. Choi, T. Kawaguchi, H. You, J. F. Mitchell, Y. Shao-Horn, and R. Comin, Anomalous antiferromagnetism in metallic $RuO_2$ determined by resonant x-ray scattering, Phys. Rev. Lett. **122**, 017202 (2019).
30. M. Hiraishi, H. Okabe, A. Koda, R. Kadono, T. Muroi, D. Hirai, and Z. Hiroi, Nonmagnetic ground state in $RuO_2$ revealed by muon spin rotation, Phys. Rev. Lett. **132**, 166702 (2024).
31. P. Keßler, L. Garcia-Gassull, A. Suter, T. Prokscha, Z. Salman, D. Khalyavin, P. Manuel, F. Orlandi, I. I. Mazin, R. Valentí, and S. Moser, Absence of magnetic order in $RuO_2$: Insights from μSR spectroscopy and neutron diffraction, npj Spintronics **2**, 50 (2024).
32. L. Kiefer, F. Wirth, A. Bertin, P. Becker, L. Bohatý, K. Schmalzl, A. Stunault, J. A. Rodríguez-Velamazan, O. Fabelo, and M. Braden, Crystal structure and absence of magnetic order in single-crystalline $RuO_2$, J. Phys.: Condens. Matter **37**, 135801 (2025).
33. Z. Wu, M. Long, H. Chen, S. Paul, H. Matsuki, O. Zheliuk, U. Zeitler, G. Li, R. Zhou, Z. Zhu, D. Graf, T. I. Weinberger, F. M. Grosche, Y. Maeno, and A. G. Eaton, Fermi surface of $RuO_2$ measured by quantum oscillations, Phys. Rev. X **15**, 031044 (2025).
34. S. G. Jeong, I. H. Choi, S. Nair, L. Buiarelli, B. Pourbahari, J. Y. Oh, B. Y. X. Lin, J. M. LeBeau, N. Bassim, D. Hirai, A. Seo, W. S. Choi, R. M. Fernandes, T. Birol, L. Zhao, J. S. Lee, and B. Jalan, Altermagnetic polar metallic phase in ultrathin epitaxially strained $RuO_2$ films, Proc. Natl. Acad. Sci. U.S.A. **123**, e2526641123 (2026).
35. D. T. Plouff, L. Scheuer, S. Shrestha, W. Wu, N. J. Parvez, S. Bhatt, X. Wang, L. Gundlach, M. B. Jungfleisch, and J. Q. Xiao, Revisiting altermagnetism in $RuO_2$: A study of laser-pulse induced charge dynamics by time-domain terahertz spectroscopy, npj Spintronics **3**, 17 (2025).
36. Y. Zhang, H. Bai, J. Dai, L. Han, C. Chen, S. Liang, Y. Cao, Y. Zhang, Q. Wang, W. Zhu, F. Pan, and C. Song, Electrical manipulation of spin splitting torque in altermagnetic $RuO_2$, Nat. Commun. **16**, 5646 (2025).
37. S. G. Jeong, S. Lee, B. Lin, Z. Yang, I. H. Choi, J. Y. Oh, S. Song, S. W. Lee, S. Nair, R. Choudhary, J. Parikh, S. Park, W. S. Choi, J. S. Lee, J. M. LeBeau, T. Low, and B. Jalan, Metallicity and anomalous Hall effect in epitaxially strained, atomically thin $RuO_2$ films, Proc. Natl. Acad. Sci. U.S.A. **122**, e2500831122 (2025).
38. I. H. Choi, S. G. Jeong, B. Jalan, and J. S. Lee, Exploring altermagnetism in $RuO_2$: From conflicting experiments to emerging consensus, Nano Converg. **13**, 1 (2026).
39. S. M. Hussain, and K. Son, Exploring altermagnetism in RuO2: A review of a new paradigm in spintronics, Physica B: Condensed Matter, 417723 (2025).
40. G. Kresse and J. Furthmüller, Efficient iterative schemes for ab initio total-energy calculations using a plane-wave basis set, Phys. Rev. B **54**, 11169 (1996).
41. P. E. Blöchl, Projector augmented-wave method, Phys. Rev. B **50**, 17953 (1994).
42. G. Kresse and D. Joubert, From ultrasoft pseudopotentials to the projector augmented-wave method, Phys. Rev. B **59**, 1758 (1999).
43. J. P. Perdew, K. Burke, and M. Ernzerhof, Generalized gradient approximation made simple, Phys. Rev. Lett. **77**, 3865 (1996).
44. P. Giannozzi, S. Baroni, N. Bonini, M. Calandra, R. Car, C. Cavazzoni, D. Ceresoli, G. L. Chiarotti, M. Cococcioni, I. Dabo, A. Dal Corso, S. de Gironcoli, S. Fabris, G. Fratesi, R. Gebauer, U. Gerstmann, C. Gougoussis, A. Kokalj, M. Lazzeri, L. Martin-Samos, N. Marzari, F. Mauri, R. Mazzarello, S. Paolini, A. Pasquarello, L. Paulatto, C. Sbraccia, S. Scandolo, G. Sclauzero, A. P. Seitsonen, A. Smogunov, P. Umari, and R. M. Wentzcovitch, QUANTUM ESPRESSO: A modular and open-source software project for quantum simulations of materials, J. Phys.: Condens. Matter **21**, 395502 (2009).
45. G. Pizzi, V. Vitale, R. Arita, S. Blügel, F. Freimuth, G. Géranton, M. Gibertini, D. Gresch, C. Johnson, T. Koretsune, J. Ibañez-Azpiroz, H. Lee, J.-M. Lihm, D. Marchand, A. Marrazzo, Y. Mokrousov, J. I. Mustafa, Y. Nohara, Y. Nomura, L. Paulatto, S. Poncé, T. Ponweiser, J. Qiao, F. Thöle, S. S. Tsirkin, M. Wierzbowska, N. Marzari, D. Vanderbilt, I. Souza, A. A. Mostofi, and J. R. Yates, Wannier90 as a community code: New features and applications, J. Phys.: Condens. Matter **32**, 165902 (2020).
46. X. Chen, J. Ren, Y. Zhu, Y. Yu, A. Zhang, P. Liu, J. Li, Y. Liu, C. Li, and Q. Liu, Enumeration and representation theory of spin space groups, Phys. Rev. X **14**, 031038 (2024).
47. S. P. Ong, W. D. Richards, A. Jain, G. Hautier, M. Kocher, S. Cholia, D. Gunter, V. L. Chevrier, K. A. Persson, and G. Ceder,

Python Materials Genomics (pymatgen): A robust, open-source python library for materials analysis, Comput. Mater. Sci. **68**, 314 (2013).

48. G. Yumnam, P. R. Raghuvanshi, J. D. Budai, L. Bocklage, D. Abernathy, Y. Cheng, A. H. Said, I. I. Mazin, H. Zhou, B. A. Frandsen, D. S. Parker, L. R. Lindsay, V. R. Cooper, M. E. Manley, and R. P. Hermann, Constraints on magnetism and correlations in $RuO_2$ from lattice dynamics and Mössbauer spectroscopy, Cell Rep. Phys. Sci. **6**, 102852 (2025).
49. F. Pawula, A. Fakih, R. Daou, S. Hébert, N. Mordvinova, O. Lebedev, D. Pelloquin, and A. Maignan, Multiband transport in $RuO_2$, Phys. Rev. B **110**, 064432 (2024).
50. A. Smolyanyuk, I. I. Mazin, L. Garcia-Gassull, and R. Valentí, Fragility of the magnetic order in the prototypical altermagnet $RuO_2$, Phys. Rev. B **109**, 134424 (2024).
51. J. Liu, J. Zhan, T. Li, J. Liu, S. Cheng, Y. Shi, L. Deng, M. Zhang, C. Li, J. Ding, Q. Jiang, M. Ye, Z. Liu, Z. Jiang, S. Wang, Q. Li, Y. Xie, Y. Wang, S. Qiao, J. Wen, Y. Sun, and D. Shen, Absence of altermagnetic spin splitting character in rutile oxide $RuO_2$, Phys. Rev. Lett. **133**, 176401 (2024).
52. T. Osumi, K. Yamauchi, S. Souma, S. Paul, A. Honma, K. Nakayama, K. Ozawa, M. Kitamura, K. Horiba, H. Kumigashira, C. Bigi, F. Bertran, T. Oguchi, T. Takahashi, Y. Maeno, and T. Sato, Spin-degenerate bulk bands and topological surface states associated with Dirac nodal lines in $RuO_2$, Phys. Rev. B **113**, 085116 (2026).
53. M. Wenzel, E. Uykur, S. Rößler, M. Schmidt, O. Janson, A. Tiwari, M. Dressel, and A. A. Tsirlin, Fermi-liquid behavior of nonaltermagnetic $RuO_2$, Phys. Rev. B **111**, L041115 (2025).
54. J. Song, C. Mu, S. Zhu, X. Zhou, W. Wu, Y. Long, J. Luo, and Z. Li, Absence of magnetic order and magnetic fluctuations in $RuO_2$, Phys. Rev. B **112**, 144444 (2025).
55. O. Fedchenko, J. Minár, A. Akashdeep, S. W. D'Souza, D. Vasilyev, O. Tkach, L. Odenbreit, Q. Nguyen, D. Kutnyakhov, N. Wind, L. Wenthaus, M. Scholz, K. Rossnagel, M. Hoesch, M. Aeschlimann, B. Stadtmüller, M. Kläui, G. Schönhense, T. Jungwirth, A. B. Hellenes, G. Jakob, L. Šmejkal, J. Sinova, and H.-J. Elmers, Observation of time-reversal symmetry breaking in the band structure of altermagnetic $RuO_2$, Sci. Adv. **10**, eadj4883 (2024).
56. Y. Zhang, S. G. Jeong, L. Buiarelli, S. Lee, Y. Guo, J. Wen, H. Li, S. Nair, I. H. Choi, Z. Ren, Z. Yue, A. Fedorov, S.-K. Mo, J. Kono, J. S. Lee, T. Low, T. Birol, R. M. Fernandes, M. Radovic, B. Jalan, and M. Yi, Observation of mirror-odd and mirror-even spin texture in ultra-thin epitaxially-strained $RuO_2$ films, arXiv:2509.16361 (2026).
57. Y.-C. Zhang, H. Bai, D.-H. Zhang, C. Chen, L. Han, S.-X. Liang, R.-Y. Chu, J.-K. Dai, M. Sawicki, F. Pan, and C. Song, Probing the Néel order in altermagnetic $RuO_2$ films via x-ray magnetic linear dichroism, Chin. Phys. Lett. **42**, 027301 (2025).
58. C. He, Z. Wen, J. Okabayashi, Y. Miura, T. Ma, T. Ohkubo, T. Seki, H. Sukegawa, and S. Mitani, Evidence for single variant in altermagnetic $RuO_2$ (101) thin films, Nat. Commun. **16**, 8235 (2025).
59. S. Wang, C. Wang, Y. Yuan, J. Li, F. Pei, D. Liu, C. Qin, J. Cao, Y. Wang, T. Wang, J. Liu, J.-E. Lee, G. Zhang, C. Klewe, C. Yu, F. Zhang, D. Song, K. Chen, W. Zhao, D. Shen, Z. Qiu, M. Yang, B. Hong, and Q. Li, Absence of magnetic order in epitaxial $RuO_2$ revealed by x-ray linear dichroism, Phys. Rev. B **113**, 024439 (2026).
60. R. Y. Chu, L. Han, Z. H. Gong, X. Z. Fu, H. Bai, S. X. Liang, C. Chen, S.-W. Cheong, Y. Y. Zhang, J. W. Liu, Y. Y. Wang, F. Pan, H. Z. Lu, and C. Song, Third-order nonlinear Hall effect in altermagnet $RuO_2$, Phys. Rev. Lett. **135**, 216703 (2025).
61. C.-T. Liao, Y.-C. Wang, Y.-C. Tien, S.-Y. Huang, and D. Qu, Separation of inverse altermagnetic spin-splitting effect from inverse spin Hall effect in $RuO_2$, Phys. Rev. Lett. **133**, 056701 (2024).
62. X. Feng, H. Bai, X. Fan, M. Guo, Z. Zhang, G. Chai, T. Wang, D. Xue, C. Song, and X. Fan, Incommensurate spin density wave in antiferromagnetic $RuO_2$ evinced by abnormal spin splitting torque, Phys. Rev. Lett. **132**, 086701 (2024).
63. H. Bai, L. Han, X. Y. Feng, Y. J. Zhou, R. X. Su, Q. Wang, L. Y. Liao, W. X. Zhu, X. Z. Chen, F. Pan, X. L. Fan, and C. Song, Observation of spin splitting torque in a collinear antiferromagnet $RuO_2$, Phys. Rev. Lett. **128**, 197202 (2022).
64. H. Chen, Z.-A. Wang, P. Qin, Z. Meng, X. Zhou, X. Wang, L. Liu, G. Zhao, Z. Duan, T. Zhang, J. Liu, D.-F. Shao, C. Jiang, and Z. Liu, Spin-splitting magnetoresistance in altermagnetic $RuO_2$ thin films, Adv. Mater. **37**, 2507764 (2025).
65. Y. Guo, J. Zhang, Z. Zhu, Y. Jiang, L. Jiang, C. Wu, J. Dong, X. Xu, W. He, B. He, Z. Huang, L. Du, G. Zhang, K. Wu, X. Han, D. Shao, G. Yu, and H. Wu, Direct and inverse spin splitting effects in altermagnetic $RuO_2$, Adv. Sci. **11**, 2400967 (2024).
66. Y. Fan, Q. Wang, W. Wang, D. Wang, Q. Huang, Z. Wang, X. Han, Y. Chen, L. Bai, S. Yan, and Y. Tian, Robust magnetic-field-free perpendicular magnetization switching by manipulating spin polarization direction in $RuO_2$/[Pt/Co/Pt] heterojunctions, ACS Nano **18**, 26350 (2024).
67. Y.-C. Wang, Z.-Y. Shen, C.-H. Lin, W.-C. Hsu, Y.-S. Chen, Y.-Y. Chin, A. Kr. Singh, W.-L. Lee, C.-T. Chen, S.-Y. Huang, and D. Qu, Absence of transport altermagnetic spin-splitting effect in $RuO_2$, Nano Lett. **26**, 2548 (2026).
68. J. Jechumtál, O. Gueckstock, K. Jasenský, Z. Kašpar, K. Olejník, M. Gaerner, G. Reiss, S. Moser, P. Kessler, G. De Luca, S. Ganguly, J. Santiso, D. Scheffler, J. Zázvorka, P. Kubaščík, H. Reichlová, E. Schmoranzerová, P. Němec, T. Jungwirth, P. Kužel, T. Kampfrath, and L. Nádvorník, Spin- to charge-current conversion in altermagnetic candidate $RuO_2$ probed by terahertz emission spectroscopy, Phys. Rev. B **113**, 054439 (2026).
69. Z. Qian, Y. Yang, S. Liu, and C. Wu, Fragile unconventional magnetism in $RuO_2$ by proximity to Landau-Pomeranchuk instability, Phys. Rev. B **111**, 174425 (2025).
70. D. T. Plouff, N. J. Parvez, S. Bhatt, X. Zhang, A. A. Aczel, J. Gaudet, and J. Q. Xiao, Competing ferromagnetic and antiferromagnetic interactions in non-altermagnetic $Ru_{1-x}Cr_xO_2$, arXiv:2607.20298 (2026).
71. A. Akashdeep, S. Krishnia, J.-H. Ha, S. An, M. Gaerner, T. Prokscha, A. Suter, G. Janka, G. Reiss, T. Kuschel, D.-S. Han, A. D. Bernardo, Z. Salman, G. Jakob, and M. Kläui, Surface-localized magnetic order in $RuO_2$ thin films revealed by low-energy muon probes, Appl. Phys. Lett. **128**, 022406 (2026).

72. D. Q. Ho, D. Q. To, R. Hu, G. W. Bryant, and A. Janotti, Symmetry-breaking induced surface magnetization in nonmagnetic $RuO_2$, Phys. Rev. Materials **9**, 094406 (2025).
73. Y. Xiang, Y. Wang, X. Li, Z. Zhao, and W. Xie, Interface-induced two-dimensional altermagnetism in $RuO_2/TiO_2$ superlattices, J. Phys.: Condens. Matter **38**, 235803 (2026).
74. P. Lu, J. S. Lee, and Q. X. Jia, Microstructural properties of $Ba_{0.6}Sr_{0.4}TiO_3/RuO_2$ multi-layers grown on MgO and YSZ by pulsed-laser deposition, Integr. Ferroelectr. **55**, 965 (2003).
75. S. Lee, S. G. Jeong, J.-P. Wang, B. Jalan, and T. Low, Strain-driven altermagnetic spin-splitting effect in $RuO_2$, Nano Lett. (2026), doi:10.1021/acs.nanolett.6c00768.
76. M. Lui, J. Drucker, A. R. King, J. P. Kotthaus, P. K. Hansma, and V. Jaccarino, Observation of antiferromagnetic resonance in epitaxial films of $MnF_2$, Phys. Rev. B **33**, 7720 (1986).
77. C. A. Ramos, D. Lederman, A. R. King, and V. Jaccarino, New antiferromagnetic insulator superlattices: Structural and magnetic characterization of $(FeF_2)_m(CoF_2)_n$, Phys. Rev. Lett. **65**, 2913 (1990).
78. J. S. Kim, Y. Jang, S. Kang, Y. Lee, K. Kim, W. Kim, W. Lee, and C. S. Hwang, Substrate-dependent growth behavior of atomic-layer-deposited zinc oxide and zinc tin oxide thin films for thin-film transistor applications, J. Phys. Chem. C **124**, 26780 (2020).
79. D. Teschner, R. Farra, L. Yao, R. Schlögl, H. Soerijanto, R. Schomäcker, T. Schmidt, L. Szentmiklósi, A. P. Amrute, C. Mondelli, J. Pérez-Ramírez, G. Novell-Leruth, and N. López, An integrated approach to Deacon chemistry on $RuO_2$-based catalysts, J. Catal. **285**, 273 (2012).
80. M. Kubovsky, Y. A. Birkhölzer, L. B. Mitrovic, H. Paik, G. R. Rossman, and D. G. Schlom, Preparation and evaluation of alexandrite, forsterite, and topaz substrates for the epitaxial growth of rutile oxides, APL Mater. **14**, 031103 (2026).
81. X. Peng, Z. Liu, S. Zhang, Y. Zhou, Y. Sun, Y. Su, C. Wu, T. Zhou, L. Liu, Y. Li, H. Wang, J. Yang, B. Chen, Y. Li, C. Xi, J. Du, Z. Jiao, Q. Wu, and M. Fang, Universal scaling behavior of transport properties in non-magnetic $RuO_2$, Commun. Mater. **6**, 177 (2025).
82. T. Qian, A. Rutherford, E. S. Choi, H. Zhou, B. Maiorov, M. Lee, and C. A. Mizzi, Determining the nature of magnetism in altermagnetic candidate $RuO_2$, Phys. Rev. Research **8**, L012072 (2026).
83. S. Paul, G. Mattoni, H. Matsuki, T. Johnson, C. Sow, S. Yonezawa, and Y. Maeno, Growth of ultra-clean oxide single crystals of the altermagnet candidate $RuO_2$, J. Cryst. Growth **673**, 128405 (2026).
84. S. Paul, A. Ikeda, H. Matsuki, G. Mattoni, J. Schmalian, K. Yamauchi, C. Sow, S. Yonezawa, and Y. Maeno, Fermi-liquid behavior and characteristic temperature-dependent susceptibility in clean single crystal $RuO_2$, Phys. Rev. B **114**, 034414 (2026).
85. Z. Lin, D. Chen, W. Lu, X. Liang, S. Feng, K. Yamagami, J. Osiecki, M. Leandersson, B. Thiagarajan, J. Liu, C. Felser, and J. Ma, Bulk band structure of $RuO_2$ measured with soft x-ray angle-resolved photoemission spectroscopy, Phys. Rev. B **111**, 134450 (2025).
86. A. Akashdeep, E. M. Ababneh, C. Schmitt, E. Galíndez-Ruales, F. Fuhrmann, T. Kuschel, M. Kläui, V. Amin, and G. Jakob, Angle-dependent magnetoresistance induced by interface-generated spin current in $RuO_2$/permalloy heterostructures, Phys. Rev. Applied **24**, 054018 (2025).
87. H. Jung, G. So, S. Noh, G.-H. Kim, J. Lee, J. Lee, S. Lee, U. Seo, D.-S. Han, Y. S. Oh, H. Jin, C. Sohn, and J.-W. Yoo, Reversible spin splitting effect in altermagnetic $RuO_2$ thin films, Nano Lett. **25**, 16985 (2025).
88. Q. Jia, S. G. Jeong, S. Lee, D. Tonini, A. Santhosh, Y. Yang, X. Li, B. Dixit, S. Liang, Y.-C. Chen, T. Low, B. Jalan, and J.-P. Wang, Epitaxial strain activates altermagnetic spin-splitting torques in $RuO_2$ (100), arXiv:2606.26023 (2026).
89. F. M. Abel, S. Bhatt, S. S. Fields, V. Sharma, D. Q. Ho, D. Wines, D. Q. To, J. C. Prestigiacomo, T. Adel, R. Torsi, M. F. Munoz, D. T. Plouff, X. Wang, B. Donovan, D. Heiman, G. M. Stephen, A. L. Friedman, G. W. Bryant, A. Janotti, M. E. Jamer, A. R. H. Walker, J. Q. Xiao, and S. P. Bennett, Probing magnetic properties of $RuO_2$ heterostructures through the ferromagnetic layer, arXiv:2508.15004 (2025).
90. T. V. A. Nguyen, H. Naganuma, T. N. H. Vu, S. DuttaGupta, Y. Saito, D. Vu, Y. Endo, S. Ikeda, and T. Endoh, Enhanced field-like torque generated from the anisotropic spin-split effect in triple-domain $RuO_2$ for energy-efficient spin-orbit torque magnetic random-access memory, Adv. Sci. **12**, 2413165 (2025).
91. Y. Zhang, H. Bai, L. Han, C. Chen, Y. Zhou, C. H. Back, F. Pan, Y. Wang, and C. Song, Simultaneous high charge-spin conversion efficiency and large spin diffusion length in altermagnetic $RuO_2$, Adv. Funct. Mater. **34**, 2313332 (2024).
92. S. Xu, Z. Zhang, F. Mahfouzi, Y. Huang, H. Cheng, B. Dai, J. Kim, D. Zhu, W. Cai, K. Shi, Z. Guo, K. Cao, B. Hong, Y. Liu, J. Yang, K. Zhang, J. Cao, F. Zhu, L. Tai, Y. Wang, S. Eimer, A. Fert, K. L. Wang, N. Kioussis, Y. Zhang, and W. Zhao, Giant spin-flop magnetoresistance in a collinear antiferromagnetic tunnel junction, Nat. Commun. **16**, 8370 (2025).
93. A. Bose, N. J. Schreiber, R. Jain, D.-F. Shao, H. P. Nair, J. Sun, X. S. Zhang, D. A. Muller, E. Y. Tsymbal, D. G. Schlom, and D. C. Ralph, Tilted spin current generated by the collinear antiferromagnet ruthenium dioxide, Nat. Electron. **5**, 267 (2022).
94. H. Jiang, L. Wang, S. Hu, W. Xie, Y. Min, J. Yang, F. Liu, Y. Lu, L. Xie, and H. Huang, Strain-induced giant enhancement of magnetism in $RuO_2$ films, Adv. Funct. Mater. **36**, e28172 (2026).
95. J. Lu, H. Chen, Z. Zhang, X. Wang, D. Xie, B. Liu, L. He, Y. Li, J. Du, Z. Wang, J. Luo, R. Zhang, Y. Xu, and X. Ruan, Surface ferrimagnetic order in $RuO_2$ film, arXiv:2604.10659 (2026).
96. S. G. Jeong, I. H. Choi, S. Lee, J. Y. Oh, S. Nair, J. H. Lee, C. Kim, A. Seo, W. S. Choi, T. Low, J. S. Lee, and B. Jalan, Anisotropic strain relaxation-induced directional ultrafast carrier dynamics in $RuO_2$ films, Sci. Adv. **11**, eadw7125 (2025).
97. T. Tschirner, P. Keßler, R. D. Gonzalez Betancourt, T. Kotte, D. Kriegner, B. Büchner, J. Dufouleur, M. Kamp, V. Jovic, L. Smejkal, J. Sinova, R. Claessen, T. Jungwirth, S. Moser, H. Reichlova, and L. Veyrat, Saturation of the anomalous Hall effect at high magnetic fields in altermagnetic $RuO_2$, APL Mater. **11**, 101103 (2023).
98. J. Song, S. H. Lee, S. Kang, D. Kim, J. H. Jeong, T. Oh, S. Lee, S. Lee, S. Lee, K.-H. Ahn, K.-W. Lee, M. Kim, T. W. Noh, B.-J. Yang, and C. Kim, Spin-orbit coupling driven magnetic response in altermagnetic $RuO_2$, Small **21**, 2407722 (2025).
99. X. He, L. Cai, Y. Liang, F. Yang, L. Yi, C. Pan, W. Wang, X. Qiu, Z. Xia, and L. Pan, Magnetic-field dependent Néel vector

perturbation and its magneto-transport properties of $RuO_2$ epitaxial films, Adv. Funct. Mater. **36**, e20771 (2026).

100. B. Z. Gregory, J. Strempfer, D. Weinstock, J. P. Ruf, Y. Sun, H. Nair, N. J. Schreiber, D. G. Schlom, K. M. Shen, and A. Singer, Strain-induced orbital-energy shift in antiferromagnetic $RuO_2$ revealed by resonant elastic x-ray scattering, Phys. Rev. B **106**, 195135 (2022).
101. H. Bai, Y. C. Zhang, Y. J. Zhou, P. Chen, C. H. Wan, L. Han, W. X. Zhu, S. X. Liang, Y. C. Su, X. F. Han, F. Pan, and C. Song, Efficient spin-to-charge conversion via altermagnetic spin splitting effect in antiferromagnet $RuO_2$, Phys. Rev. Lett. **130**, 216701 (2023).
102. C.-E. Boman, J. Danielsen, A. Haaland, B. Jerslev, C. E. Schäffer, E. Sunde, and N. A. Sørensen, Refinement of the crystal structure of ruthenium dioxide, Acta Chem. Scand. **24**, 116 (1970).
103. J. Haines, J. M. Léger, O. Schulte, and S. Hull, Neutron diffraction study of the ambient-pressure, rutile-type and the high-pressure, $CaCl_2$-type phases of ruthenium dioxide, Acta Crystallogr. B **53**, 880 (1997).
104. See Supplemental Material for the effects of Hubbard $U$ and carrier doping on bulk and strained $RuO_2$, nonmagnetic band structures and Fermi surfaces, band-resolved susceptibility contributions, altermagnetic band structures, RPA susceptibilities, virtual-crystal and supercell calculations of chemical doping, and structural views of $RuO_2/TiO_2$ heterostructures.
105. D. Wickramaratne, M. Currie, S. S. Fields, C. D. Cress, and S. P. Bennett, Effects of altermagnetic order, strain, and doping in $RuO_2$, J. Mater. Chem. C **14**, 1587 (2026).
106. T. Moriya and Y. Takahashi, Spin fluctuations in itinerant electron magnetism, J. Phys. Colloq. **39**, C6-1466 (1978).
107. S. Das and I. I. Mazin, Quantitative assessment of the role of spin fluctuations in 2D Ising superconductor $NbSe_2$, Comput. Mater. Sci. **200**, 110758 (2021).
108. I. Park, D. Kim, J. Hong, B. Goh, and B. G. Jang, Correlation-driven tunability of altermagnetism in $RuO_2$, arXiv:2605.13559 (2026).
109. M. Uchida, T. Nomoto, M. Musashi, R. Arita, and M. Kawasaki, Superconductivity in uniquely strained $RuO_2$ films, Phys. Rev. Lett. **125**, 147001 (2020).
110. K. Robinson, G. V. Gibbs, and P. H. Ribbe, Quadratic elongation: A quantitative measure of distortion in coordination polyhedra, Science **172**, 567 (1971).
111. M. E. Fleet, Distortion parameters for coordination polyhedra, Mineral. Mag. **40**, 531 (1976).
112. Z. Fatima, D. Oka, and T. Fukumura, Systematic application of extremely large strain to rutile-type $RuO_2$ (100) epitaxial thin films on substrates with large lattice mismatches, Cryst. Growth Des. **21**, 4083 (2021).
113. Q. Wang, D. Gilmer, Y. Fan, A. Franciosi, D. F. Evans, W. L. Gladfelter, and X. F. Zhang, Microstructure of ruthenium dioxide films grown on $\alpha$–$Al_2O_3$ (0001), $\alpha$–$Al_2O_3$ (1102), and $SrTiO_3$ (100) using reactive sputtering, J. Mater. Res. **12**, 984 (1997).
114. M. Roig, A. Kreisel, Y. Yu, B. M. Andersen, and D. F. Agterberg, Minimal models for altermagnetism, Phys. Rev. B **110**, 144412 (2024).
115. D. Csontosová, K.-H. Ahn, and J. Kuneš, Incipient magnetic instability in $RuO_2$ with random phase approximation, Low Temp. Phys. **52**, 1063 (2026).
116. N. Heinsdorf, Altermagnetic instabilities from quantum geometry, Phys. Rev. B **111**, 174407 (2025).
117. S. Graser, T. A. Maier, P. J. Hirschfeld, and D. J. Scalapino, Near-degeneracy of several pairing channels in multiorbital models for the Fe pnictides, New J. Phys. **11**, 025016 (2009).
118. C. Lu and L.-H. Hu, Van Hove singularity-induced multiple magnetic transitions in multi-orbital systems, npj Spintronics (2026), doi:10.1038/s44306-026-00156-w.
119. A. Smolyanyuk, L. Šmejkal, and I. I. Mazin, Origin of the anomalous Hall effect in Cr-doped $RuO_2$, Phys. Rev. B **111**, 064406 (2025).
120. M. Meinert, A meta-GGA perspective on the altermagnetism of $RuO_2$, Phys. Rev. Materials **10**, 054411 (2026).
121. Y. F. Hou, S. Liu, W. Fen, J. Lu, X. Chen, G. B. Liu, and P. Zhang, Nonmagnetic-magnetic transitions in rutile $RuO_2$, arXiv:2604.14764 (2026).
122. W. Liu, X. Feng, J. Zhao, P. Zhang, D. Guo, D. Xue, M. Si, Z. Li, F. Guo, and X. Fan, Strain-induced magnetic order in $RuO_2$ from first-principles calculations, Phys. Rev. B **114**, 144417 (2026).
123. C. Tang, F. Li, and L. Tang, Strain-induced magnetization in $RuO_2$, Phys. Rev. B **114**, 074412 (2026).
124. P. Keßler, T. Waldsauer, V. Jovic, M. Kamp, M. Schmitt, M. Sing, R. Claessen, and S. Moser, Epitaxial $RuO_2$ and $IrO_2$ films by pulsed laser deposition on $TiO_2$ (110), arXiv:2405.12878 (2026).
125. M. Wang, K. Tanaka, S. Sakai, Z. Wang, K. Deng, Y. Lyu, C. Li, D. Tian, S. Shen, N. Ogawa, N. Kanazawa, P. Yu, R. Arita, and F. Kagawa, Emergent zero-field anomalous Hall effect in a reconstructed rutile antiferromagnetic metal, Nat. Commun. **14**, 8240 (2023).
126. S. G. Jeong, A. Santhosh, S. Lee, S. Sheikh, U. M. Jayathilake, T.-L. Lee, T. Low, A. X. Gray, S. A. Chambers, and B. Jalan, Disentangling strain and $Ti^{3+}$ contributions to the anomalous Hall effect in epitaxial $RuO_2$ films, arXiv:2609.18059 (2026).

SUPPLEMENTAL MATERIAL

# Conditions for Altermagnetism in $RuO_2$

Himanshu Mavani, Zhonglin He, and Evgeny Y. Tsymbal

*Department of Physics and Astronomy & Nebraska Center for Materials and Nanoscience, University of Nebraska, Lincoln, Nebraska 68588-0299, USA*

**SECTION I:** The magnetic ground state of bulk $RuO_2$ as function of Hubbard U and number of electrons.

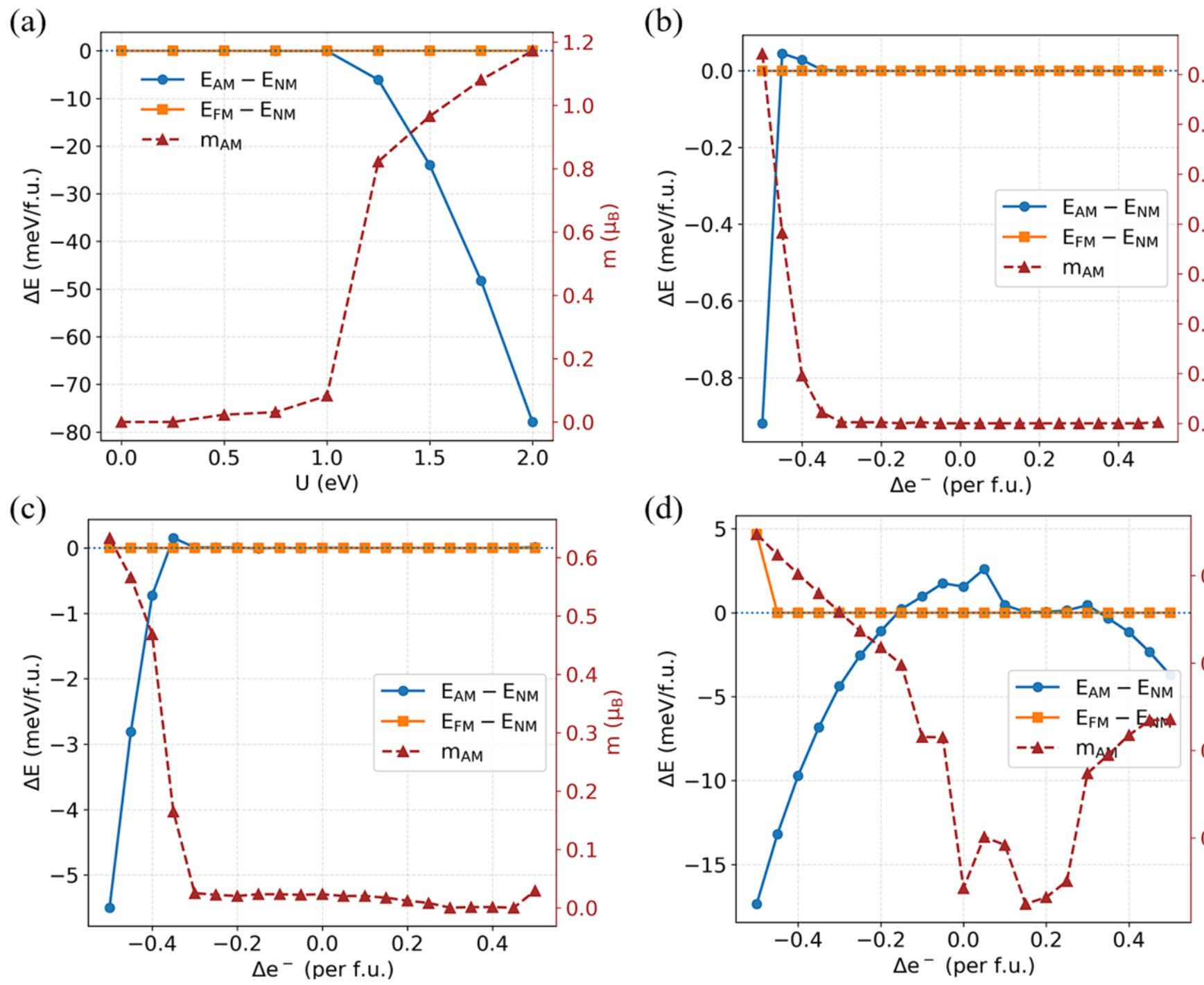

**Figure S1:** (a) The magnetic ground state of bulk $RuO_2$ as function of Hubbard U. (b-d) The magnetic ground state of bulk $RuO_2$ as function of doping with Hubbard U=0 eV (b), U=0.5 eV (c), and U=1 eV (d). Negative (positive) sign of $\Delta e^-$ denotes the hole (electron) doping.

**SECTION II:** The magnetic ground state of $RuO_2$ under epitaxially strain condition of $TiO_2$(110) configuration as function of Hubbard U and number of electrons.

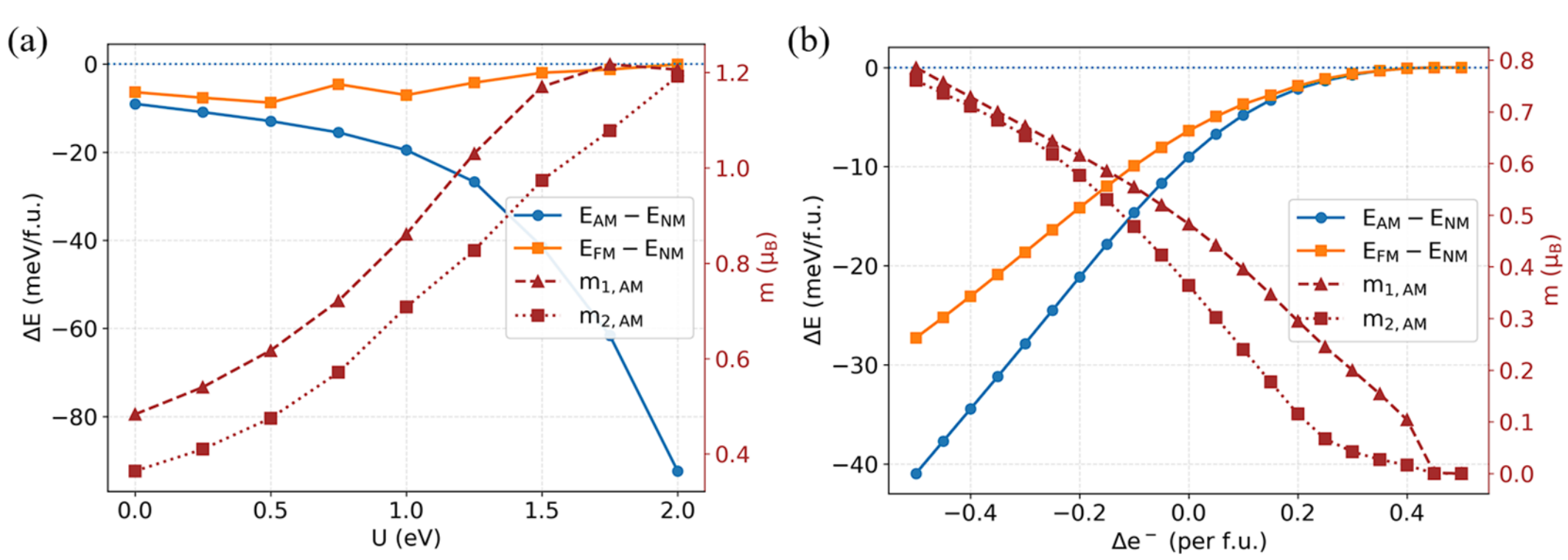

**Figure S2:** (a) The magnetic ground state of $RuO_2$ epitaxially strained to $TiO_2$(110) as function of Hubbard U. (b) The magnetic ground state of bulk $RuO_2$ as function of doping with U = 0. Negative (positive) sign of $\Delta e^-$ denotes the hole (electron) doping. The magnetic moments of two inequivalent Ru sublattices are denoted as $m_1$ and $m_2$.

**SECTION III:** Altermagnetic band structures of strained bulk systems that exhibit an altermagnetic ground state without a Hubbard (U) correction.

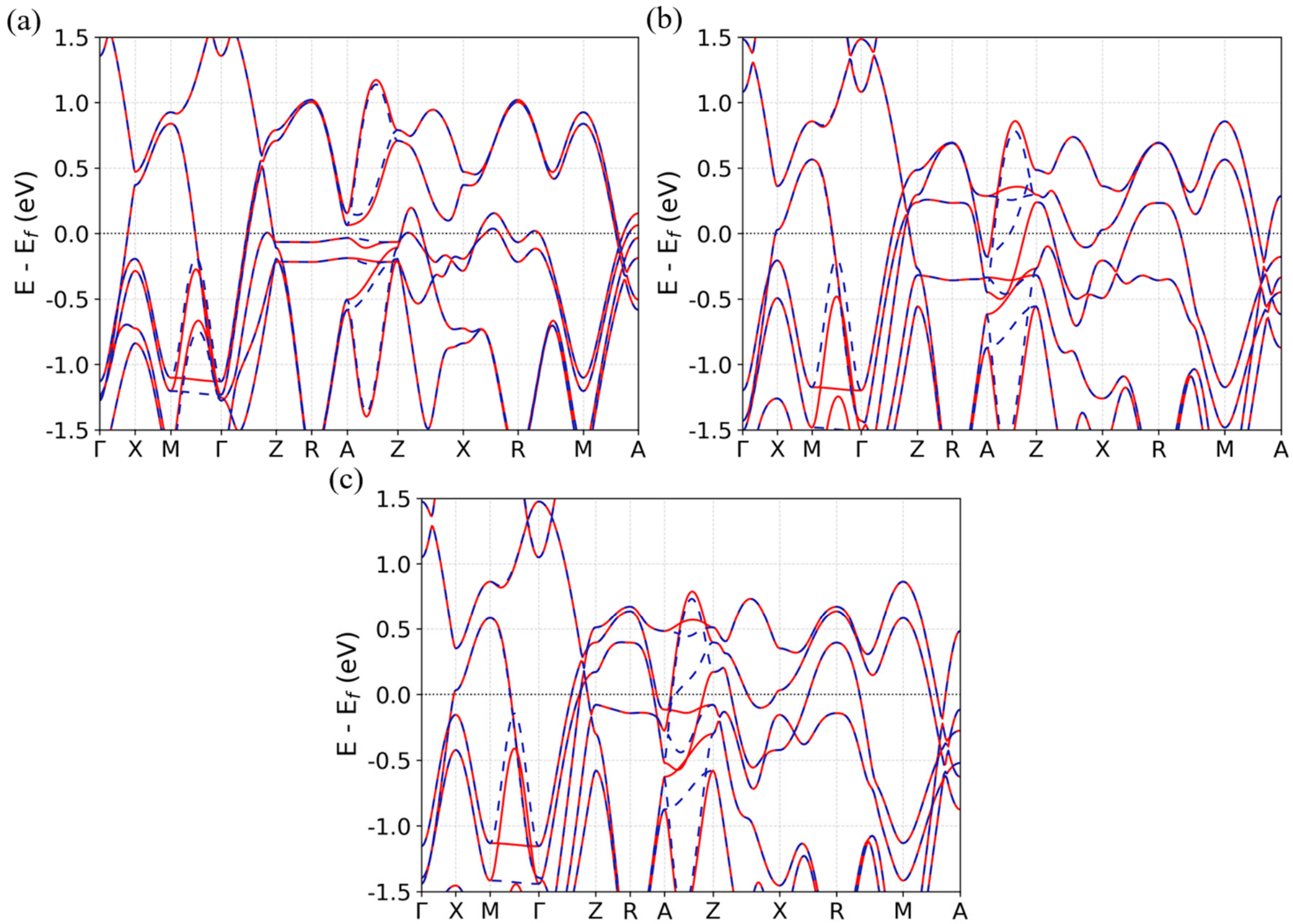

**Figure S3:** (a) Altermagnetic band structure of (a) $RuO_2$(001)/$MgF_2$(001), (b) $RuO_2$(001)/$ZnF_2$(001), and (c) $RuO_2$(001)/$SnO_2$(001).

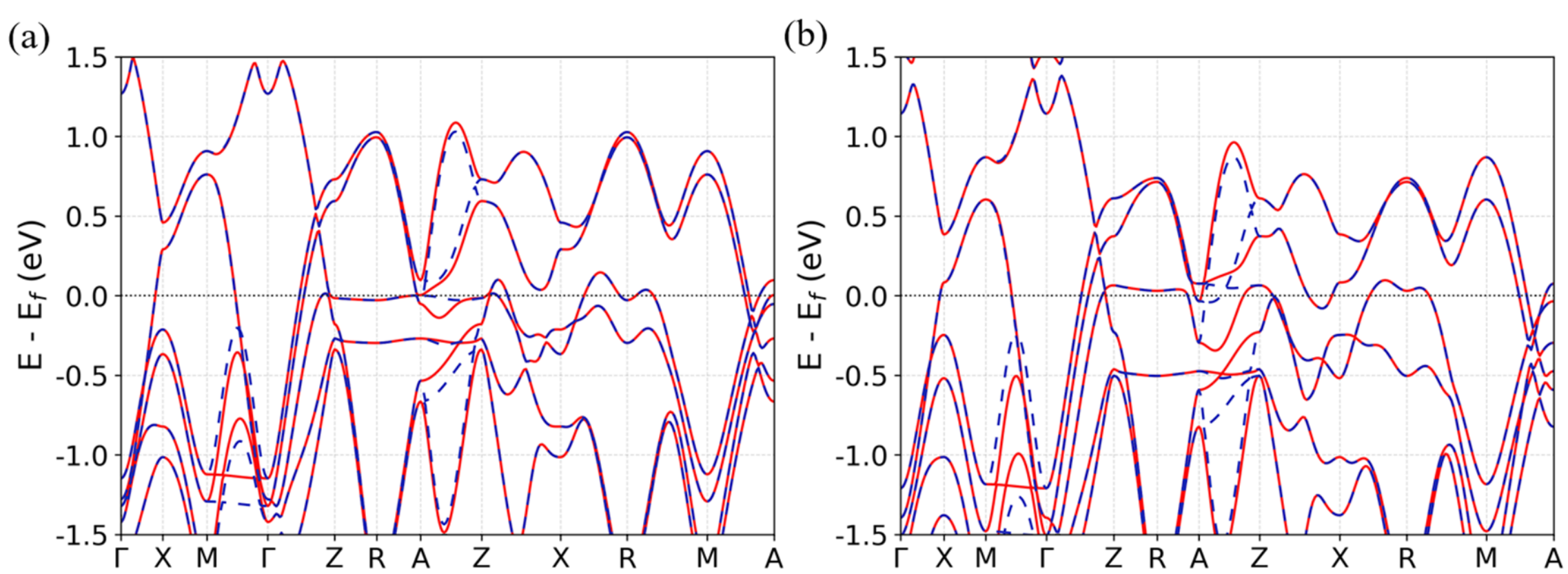

**Figure S4:** (a) Altermagnetic band structure of (a) $RuO_2$(100)/$Mg_2SiO_4$(010), and (b) $RuO_2$(100)/$TiO_2$(100).

**SECTION IV:** Nonmagnetic band structures and Fermi surfaces of the $RuO_2$ under the bulk-strained conditions imposed by the different substrates and orientations.

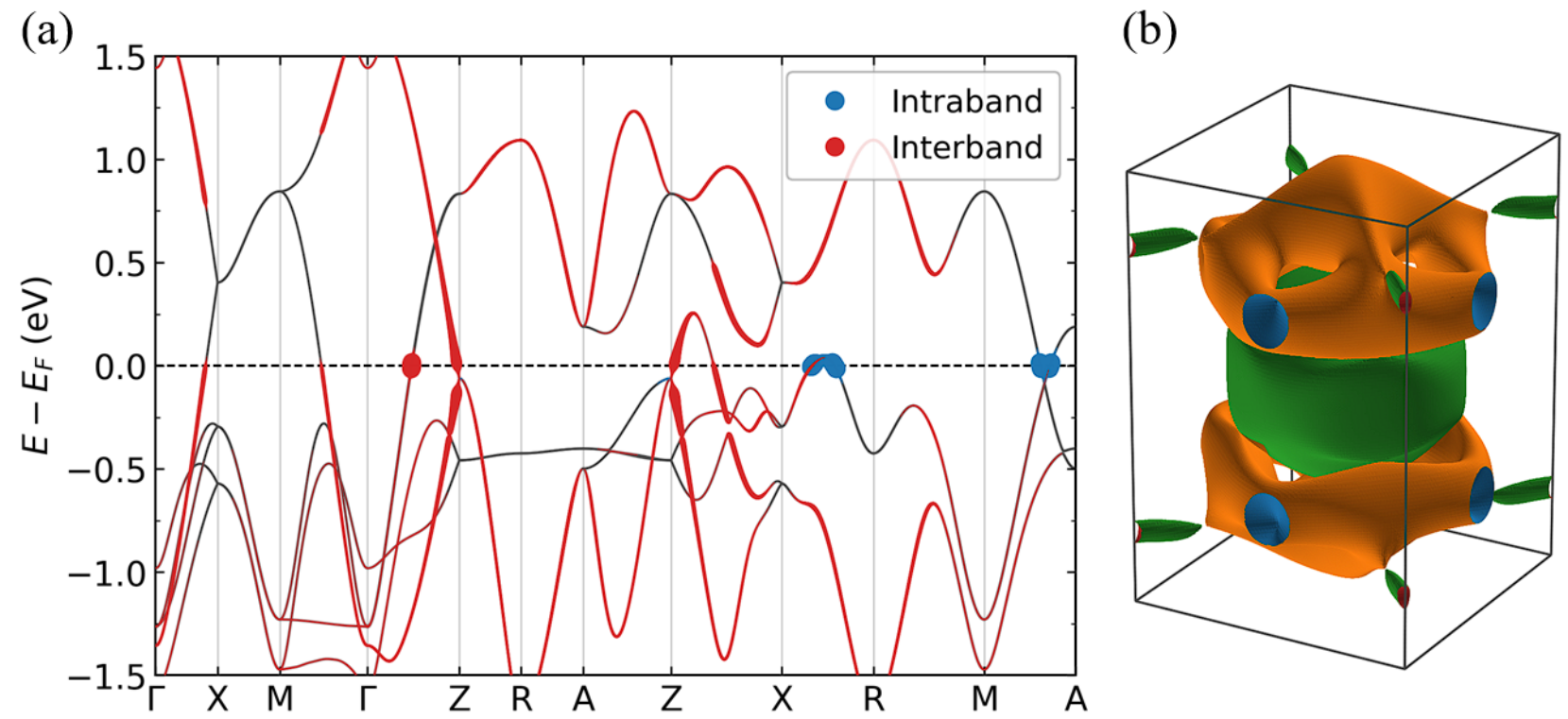

**Figure S5:** (a) Band-resolved intraband and interband contributions to $\chi_0^{\mathrm{AM}}(\Gamma)$ for strained-bulk configuration of $RuO_2$(100)/$SrTiO_3$(100). (b) Corresponding Fermi surfaces of nonmagnetic state.

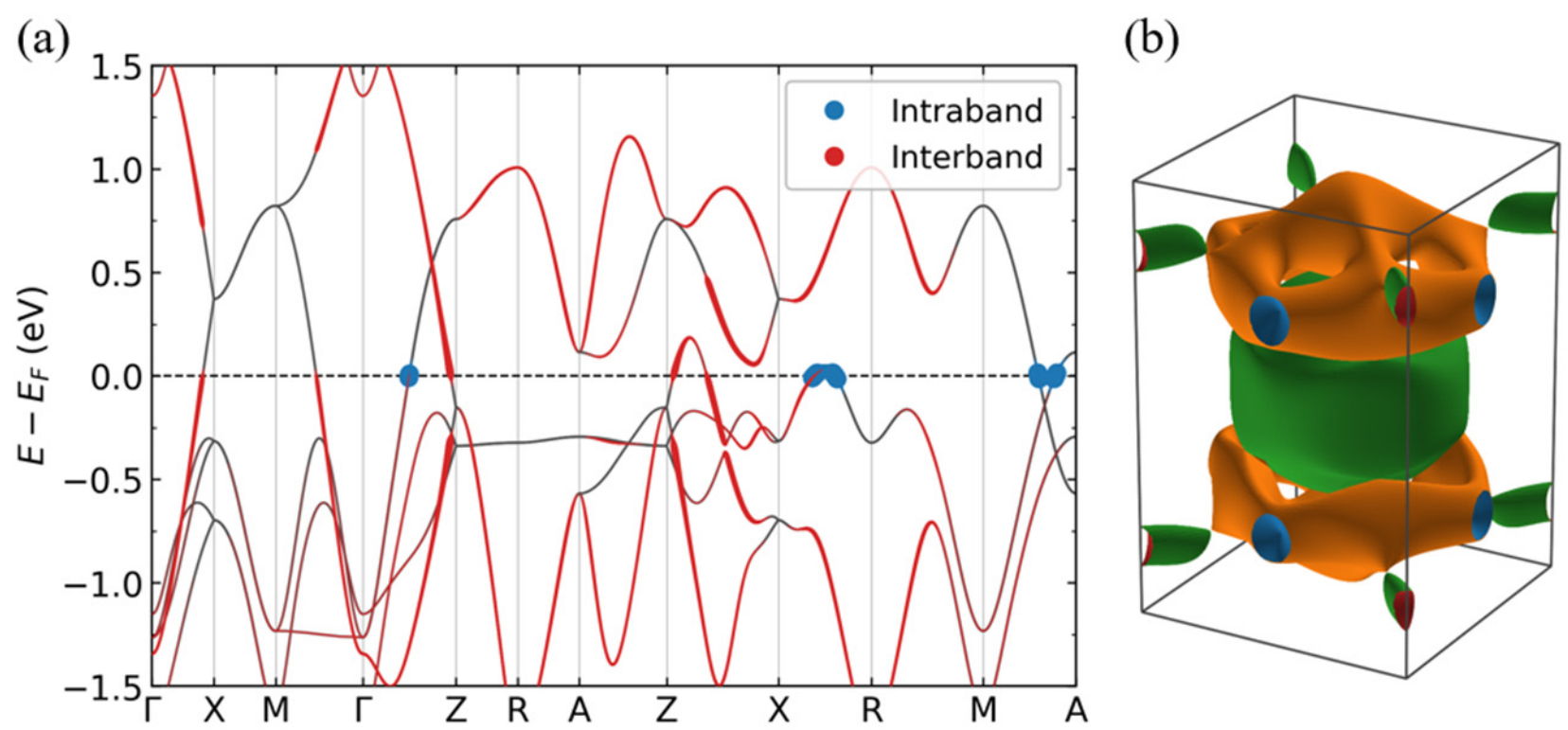

**Figure S6:** (a) Band-resolved intraband and interband contributions to $\chi_0^{\mathrm{AM}}(\Gamma)$ for strained-bulk configuration of $RuO_2$(001)/$TiO_2$(001). (b) Corresponding Fermi surfaces of nonmagnetic state.

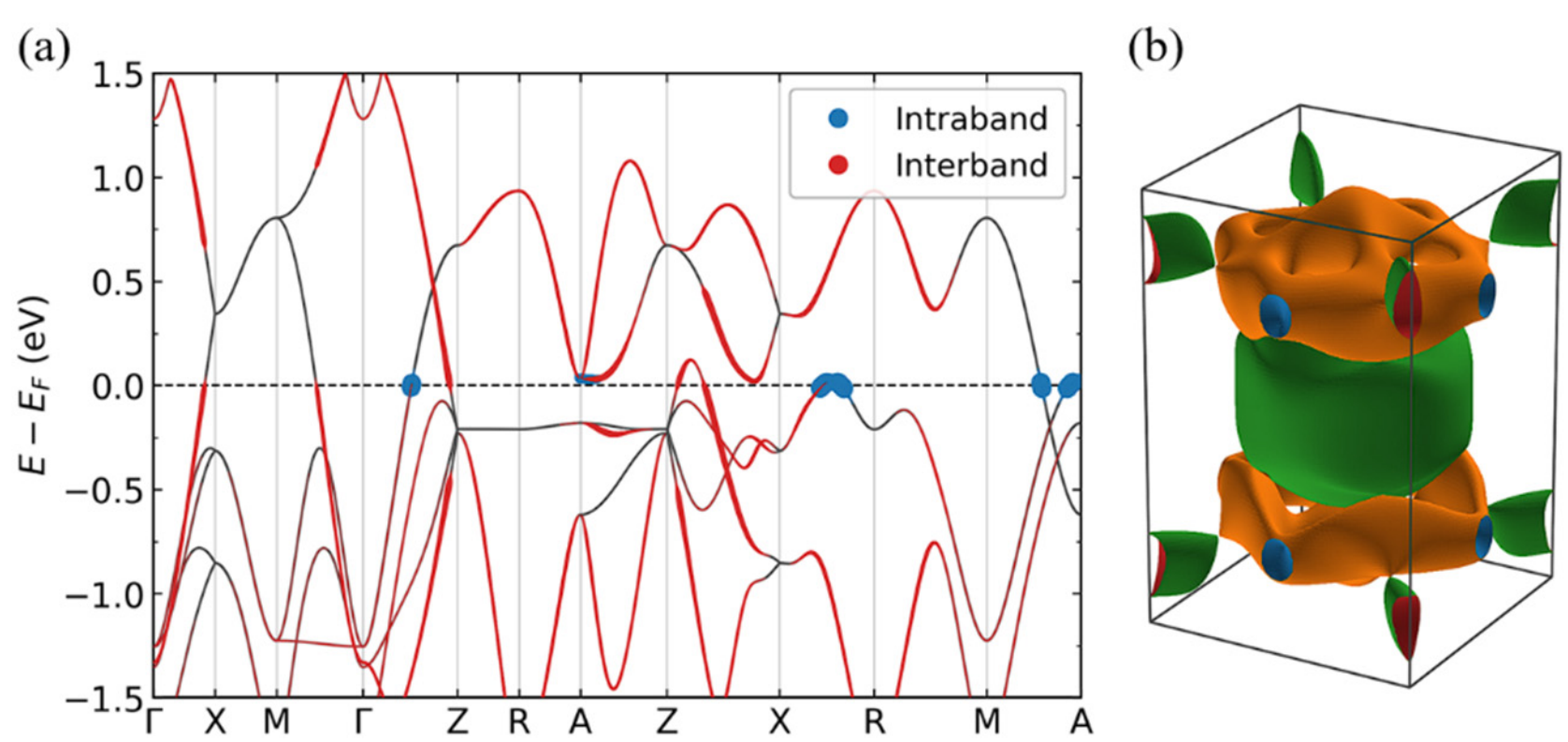

**Figure S7:** (a) Band-resolved intraband and interband contributions to $\chi_0^{\mathrm{AM}}(\Gamma)$ for strained-bulk configuration of $RuO_2$(001)/$MgF_2$(001). (b) Corresponding Fermi surfaces of nonmagnetic state.

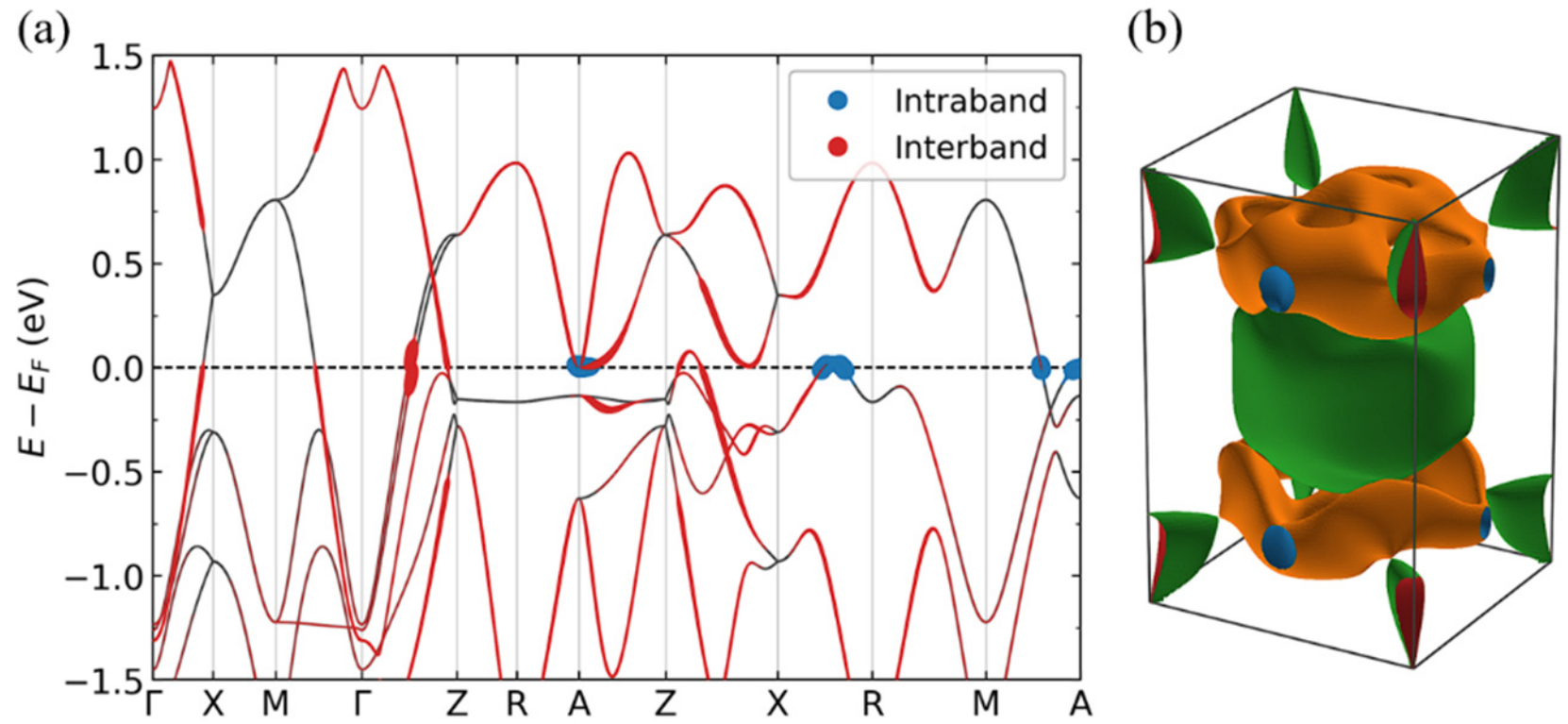

**Figure S8:** (a) Band-resolved intraband and interband contributions to $\chi_0^{\mathrm{AM}}(\Gamma)$ for strained-bulk configuration of $RuO_2(100)/Mg_2SiO_4(010)$. (b) Corresponding Fermi surfaces of nonmagnetic state.

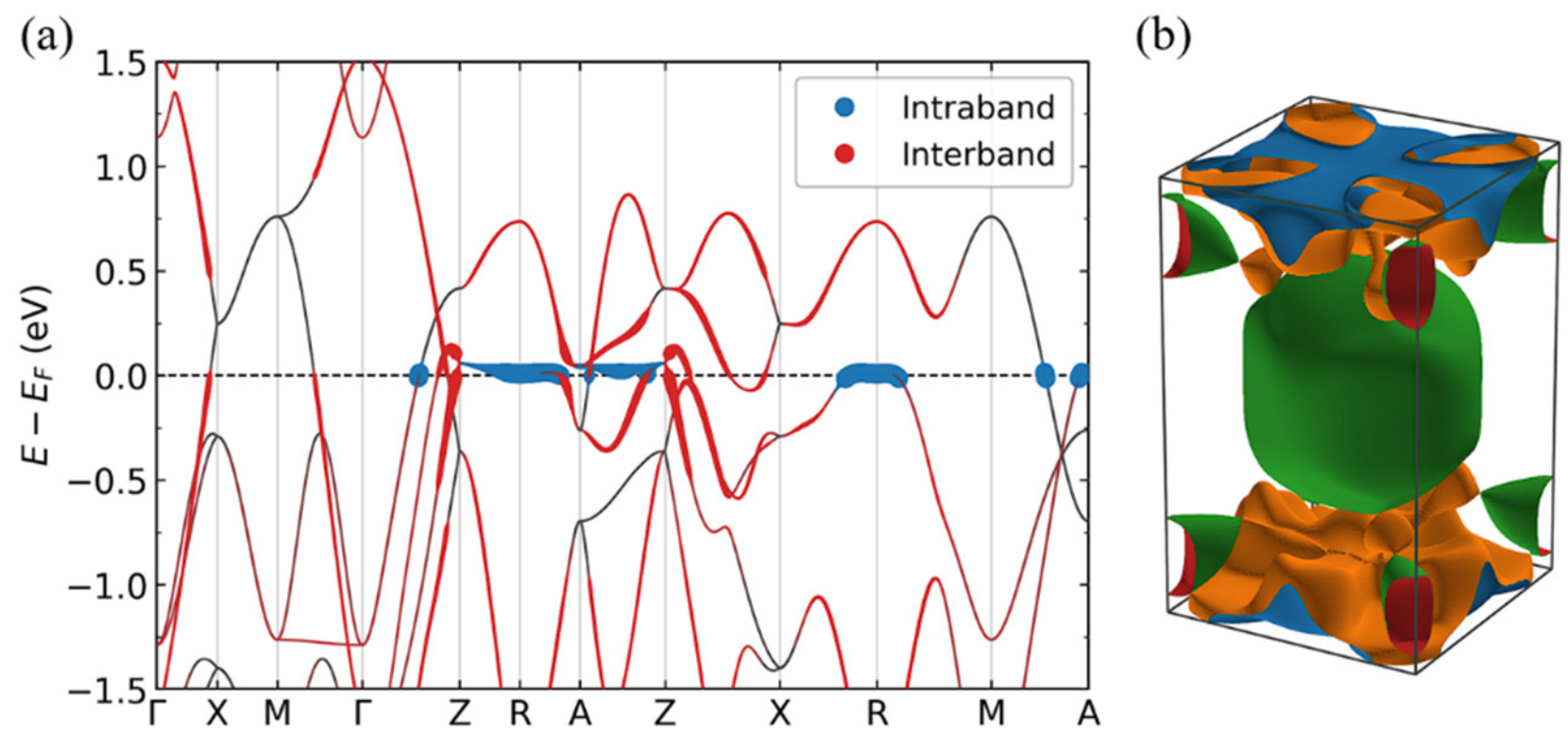

**Figure S9:** (a) Band-resolved intraband and interband contributions to $\chi_0^{\mathrm{AM}}(\Gamma)$ for strained-bulk configuration of $RuO_2(001)/ZnF_2(001)$. (b) Corresponding Fermi surfaces of nonmagnetic state.

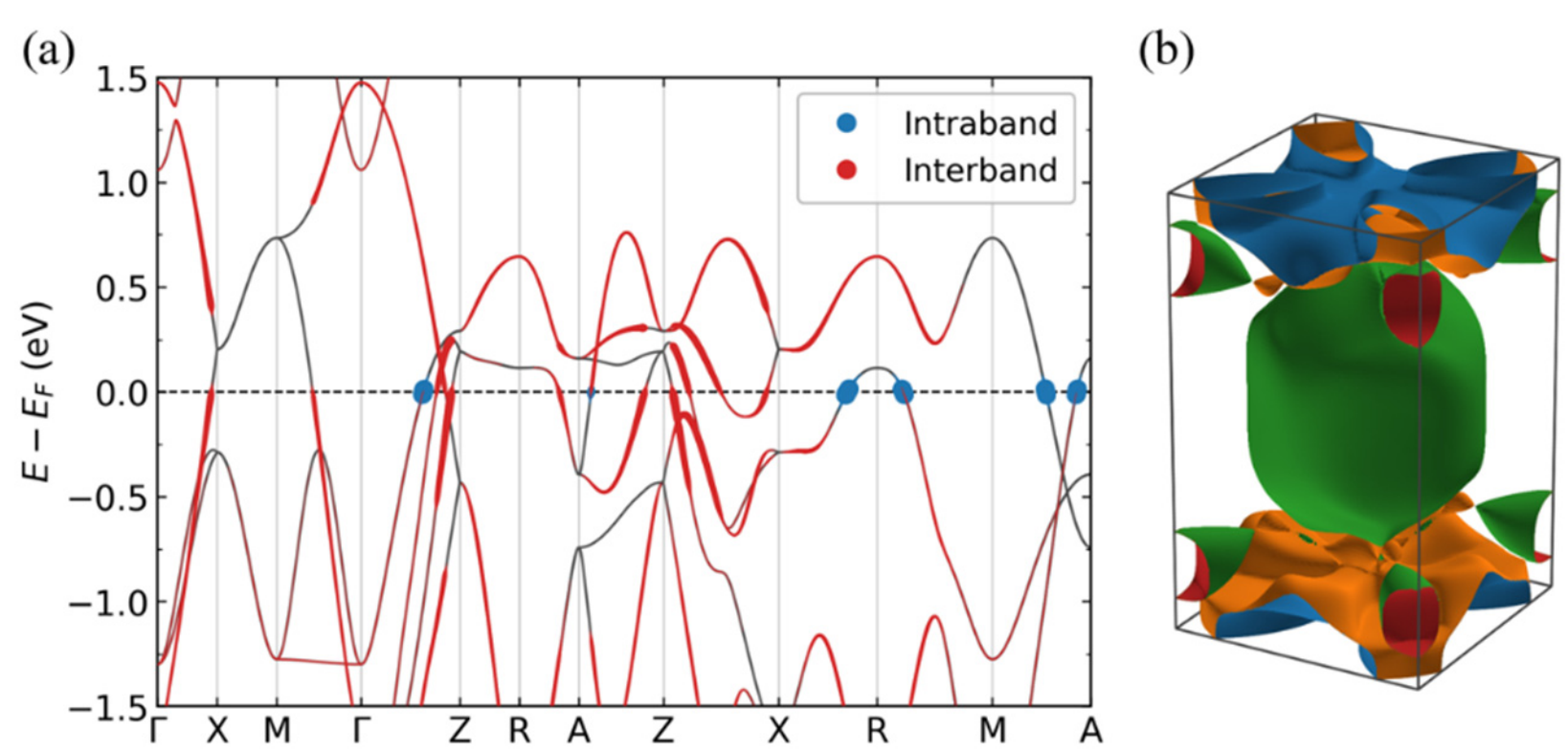

**Figure S10:** (a) Band-resolved intraband and interband contributions to $\chi_0^{\mathrm{AM}}(\Gamma)$ for strained-bulk configuration of $RuO_2(001)/SnO_2(001)$. (b) Corresponding Fermi surfaces of nonmagnetic state.

**SECTION V:** The RPA susceptibilities of $RuO_2$ under the epitaxial strain conditions of the substrates.

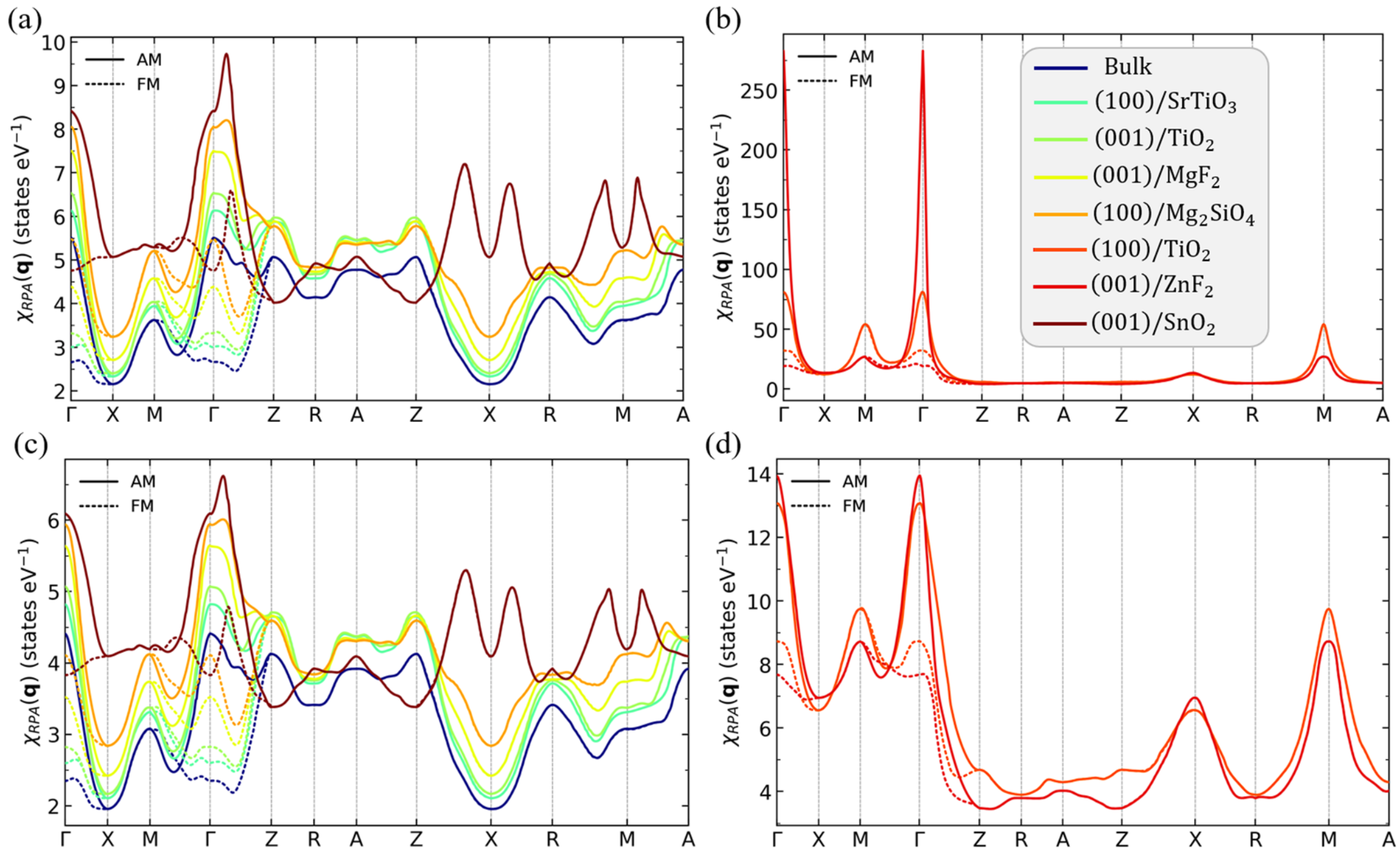

**Figure S11:** (a, b) The RPA susceptibilities with $U = 0.75\ eV$ and $J/U = 0.15$ of $RuO_2$ under the epitaxial strain conditions of the substrates. (c, d) Same with $U = 0.5\ eV$ and $J/U = 0.2$.

**SECTION VI:** The virtual crystal approximation (VCA) of doping in strained $RuO_2$.

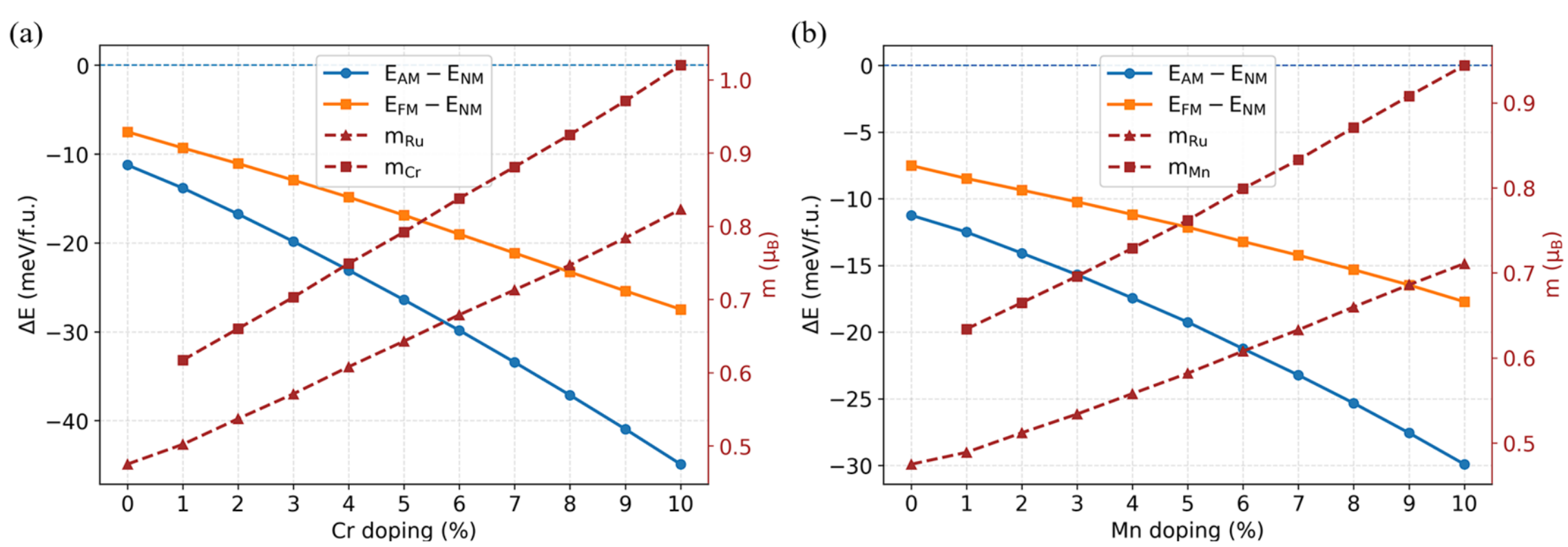

**Figure S12:** (a, b) The VCA calculation of Cr doping (a) and Mn doping (b) in $RuO_2$ epitaxially strained to $TiO_2$(100).

**SECTION VII:** The supercell calculation of Mn doping in strained $RuO_2$.

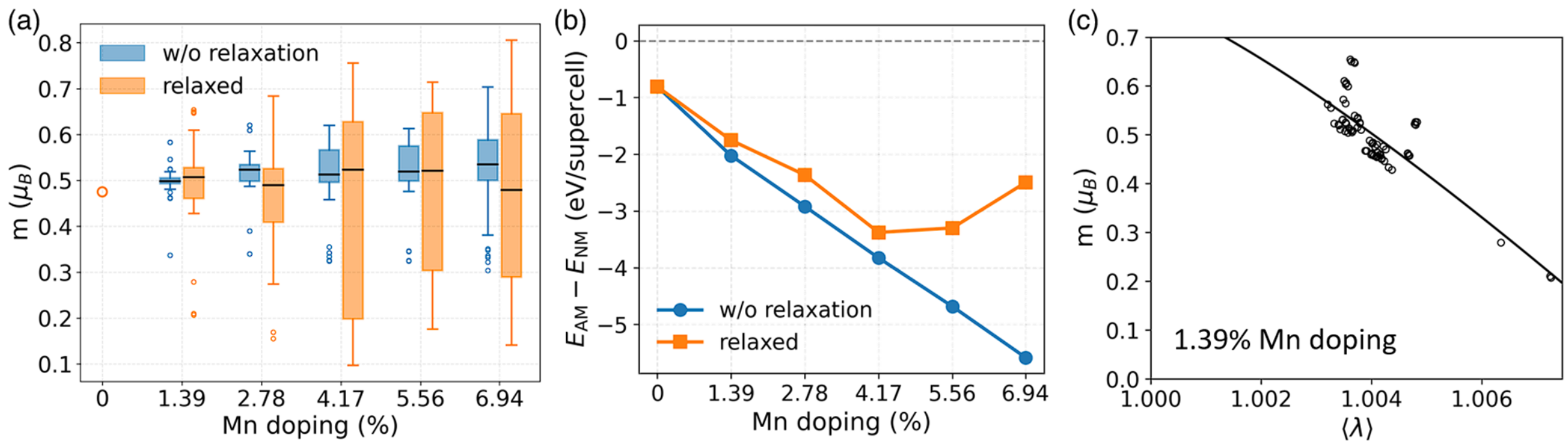

**Figure S13:** The Mn doping dependence of the magnetism for $RuO_2$ strained to epitaxial conditions of $TiO_2$(100). (a,b) Distributions of the Ru magnetic moments and the energy difference $E_{AM} - E_{NM}$ as functions of the Mn doping concentration in strained $RuO_2$ supercells. (c) Ru magnetic moments as functions of the mean quadratic elongation of the $RuO_6$ octahedra at a 1.39% Mn-doping. The Mn atoms develop large local moments of approximately 2.6 $\mu_B$.

**SECTION VIII:** The side view of $RuO_2/TiO_2$ heterostructure.

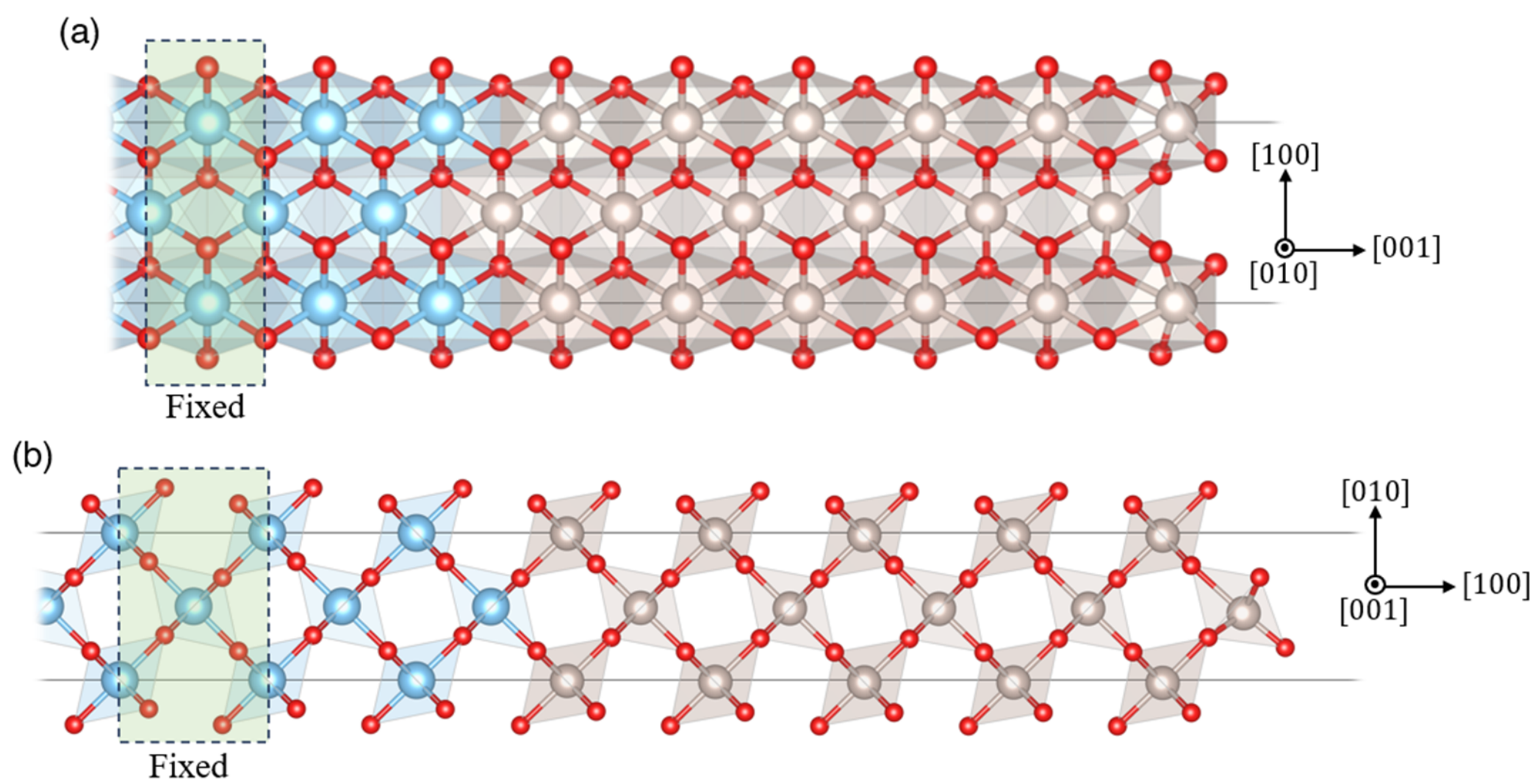

**Figure S14:** (a, b) The side view of $RuO_2/TiO_2$ heterostructure of (001) (a) and (100) (b) orientations.